\documentclass[twocolumn,showpacs,preprintnumbers,amsmath,amssymb]{revtex4}

\usepackage{hyperref}

\usepackage{graphicx,epsfig}

\newcommand{\apref}[1]{{\ref{#1}}}

\newcommand{\be}{\begin{equation}}
\newcommand{\ee}{\end{equation}}
\newcommand{\bea}{\begin{eqnarray}}
\newcommand{\eea}{\end{eqnarray}}
\newcommand{\beaa}{\begin{eqnarray*}}
\newcommand{\eeaa}{\end{eqnarray*}}

\newcommand{\ba}{\begin{array}}
\newcommand{\ea}{\end{array}}
\newcommand{\bi}{\begin{itemize}}
\newcommand{\ei}{\end{itemize}}
\newcommand{\ben}{\begin{enumerate}}
\newcommand{\een}{\end{enumerate}}

\newcommand{\bra}{\langle}
\newcommand{\ket}{\rangle}
\newcommand{\ra}{\rightarrow}
\newcommand{\lra}{\longrightarrow}

\newcommand{\lb}{\label}
\newcommand{\g}{\gamma}
\newcommand{\G}{\Gamma}

\newcommand{\al}{\alpha}

\newcommand{\p}{\partial}
\newcommand{\dl}{\delta}
\newcommand{\Dl}{\Delta}
\newcommand{\ld}{\lambda}

\newcommand{\Om}{\Omega}

\newcommand{\sm}{\sigma}

\newcommand{\mcE}{{\mathcal{E}}}

\newcommand{\OO}{{\mathcal{O}}}

\newcommand{\vx}{{\bf x}}

\begin{document}







\title{Pulsars versus Dark Matter\\
Interpretation of ATIC/PAMELA}
\author{Dmitry Malyshev}
 \email{dm137@nyu.edu}
 \altaffiliation{On leave of absence from ITEP, 
 B. Cheremushkinskaya 25, Moscow, Russia} 
\author{Ilias Cholis}
 \email{ijc219@nyu.edu}
\author{Joseph Gelfand}
 \email{jg168@astro.physics.nyu.edu}
\affiliation{%
Center for Cosmology and Particle Physics\\
 4 Washington Place, Meyer Hall of Physics, NYU, New York, NY 10003
}%

\date{\today}

\begin{abstract}

In this paper, we study the flux of electrons and positrons injected
by pulsars and by annihilating or decaying dark matter in the context
of recent ATIC, {\it PAMELA}, {\it Fermi}, and HESS data.  We review
the flux from a single pulsar and derive the flux from a distribution
of pulsars.  We point out that the particle acceleration in the pulsar
magnetosphere is insufficient to explain the observed excess of
electrons and positrons with energy $E\sim 1$ TeV and one has to take
into account an additional acceleration of electrons at the
termination shock between the pulsar and its wind nebula.  We show
that at energies less than a few hundred GeV, the flux from a
continuous distribution of pulsars provides a good approximation to
the expected flux from pulsars in the Australia Telescope National Facility
(ATNF) catalog.  
At higher energies, we demonstrate that the electron/positron flux measured at
the Earth will be dominated by a few young nearby pulsars, and
therefore the spectrum would contain bumplike features.  We argue
that the presence of such features at high energies would strongly
suggest a pulsar origin of the anomalous contribution to electron and
positron fluxes.  The absence of features either points to a dark
matter origin or constrains pulsar models in such a way that the
fluctuations are suppressed.
Also we derive that the features can be partially smeared due to
spatial variation of the energy losses during propagation.
\end{abstract}

\pacs{97.60.Gb, 
95.35.+d, 
96.50.S-, 
98.70.Sa 
}

\maketitle


\tableofcontents


\section{Introduction}
\lb{sec:Intro}

The nature of dark matter (DM) remains one of the most interesting
problems in cosmology and astrophysics.  Recently, several cosmic ray
experiments reported higher than expected fluxes of electrons and
positrons at energies between 10 GeV and 1 TeV \cite{:2008zzr,
Adriani:2008zr, Collaboration:2008aaa, Aharonian:2009ah, Abdo:2009zk}.
One interpretation is that this excess is a result of DM annihilation
or decay \cite{Cholis:2008vb, Bergstrom:2008gr, Cirelli:2008pk,
Cholis:2008hb, ArkaniHamed:2008qn, Yin:2008bs, Hamaguchi:2008rv,
Chen:2008fx}.  Standard astrophysical sources, such as pulsars
\cite{1995A&A...294L..41A, Kobayashi:2003kp, Hooper:2008kg, Yuksel:2008rf,
Profumo:2008ms, Ioka:2008cv}, are also a viable possibility.  The main
purpose of this work is to identify an observational signature which
might differentiate between these two models.

Pulsars are known to produce and accelerate electrons and positrons in
their magnetosphere \cite{Rees:1974nr}.  
However, as we will argue below, this
acceleration is insufficient to explain the observed excess.  
Moreover, the spectrum of particles in the magnetosphere is 
further modified at the termination shock between the
pulsar and the interstellar medium (ISM) or
the supernova remnant (SNR) 
which may contain the pulsar wind nebula (PWN).
It is important to emphasize that 
termination shocks are observed
around middle-aged pulsars (e.g., Geminga
\cite{2003Sci...301.1345C, Pavlov:2005wc}, PSR J1747-2958
\cite{Gaensler:2003iy}) and not just young pulsars like the Crab.
Most young energetic pulsars are also surrounded by a PWN 
(for a review see, e.g., \cite{Gaensler:2006ua})
which is powered by the continuous emission from the pulsar and 
remains inside the shell or envelope produced by the initial explosion
(a discussion of various regions surrounding a pulsar can be found in, e.g.,
\cite{Kennel:1984vf}).
Inside a PWN, the electrons and positrons are confined by the
nebula's magnetic field
for a long period of time before escaping into the ISM.  
Since a pulsar loses the vast majority of the spin-down energy while its PWN still exists, 
we assume that most of the electrons and positrons injected by pulsars
spend a significant amount of time inside a PWN before reaching the ISM.
This has two main consequences: 
\bi
\item
The spectrum of electrons and positrons injected by a pulsar into the
surrounding ISM is not the spectrum of particles inside the
magnetosphere (as assumed by many authors, e.g \cite{Hall:2008qu,
Hooper:2008kg, Profumo:2008ms}), but the spectrum of particles that
escape the PWN into the surrounding ISM.  In this paper, we assume
that this is the same as the electron/positron spectrum inside the PWN
when it is disrupted, which we estimate using the observed broadband
spectrum of these objects.
\item
Since the lifetime of a PWN ($t \ll 100$ kyr; \cite{Gaensler:2006ua}) is
generally much smaller than the typical propagation time ($t > 100$ kyr), 
the electrons observed at the Earth come from
PWNe that no longer exist, whereas electrons inside existing PWNe
cannot reach us.  
Since the variability of PWNe properties is very large, 
we cannot predict the electron flux from a pulsar that has
already lost its PWN, even if we fully know its current
properties (e.g age, position, spin-down luminosity).  
\ei

Therefore, the best one can do is to use the currently
observed PWNe to derive a statistical distribution of their properties.  
This distribution can be used to find either the
average flux of electrons expected from pulsars or, by assigning the
PWNe random properties according to the distribution, the typical
electron flux observed on Earth from all pulsars.

The observed spectrum on Earth of electrons and positrons injected by
pulsars is also strongly dependent on propagation effects.  
In particular, the observed cutoff in the flux of electrons from a pulsar can be much 
smaller than the injection cutoff due to energy losses (``cooling") during propagation.
We define the cooling break, $E_{\rm br}(t)$, as the maximal energy
electrons can have after propagating for time $t$.  Since -- as stated
above -- the typical electron propagation time is much larger than the
lifetime of a PWN, we can assume that a pulsar is a delta-function
source $Q \sim \dl(t)$ and the propagation time for electrons from
this pulsar can be estimated by the pulsar's age $t$.  
If the cooling break is at a lower energy than
the injection cutoff from
a PWN, $E_{\rm br}(t) < E_{\rm inj}$, 
then the observed break is the cooling break which
depends on the age of the pulsar but is independent of the injection cutoff.  
Since the cooling break is much steeper than the injection
break, the existence of several pulsars with \hbox{$E_{\rm br} \ll
E_{\rm inj}$} sufficiently close to the Earth such that the
propagation time of electrons they inject into the ISM is less than
their age will result in a sequence of steps or bumps in the spectrum.
At high energies, only a few pulsars will satisfy this criterion, so
these steps are expected to be well separated and therefore
observable.  At lower energies, we expect many pulsars to contribute,
causing these steps to be averaged together and resulting in a smooth
spectrum.  We argue that the presence of significant steps, or bumps,
in the electron spectrum at high energies would strongly suggest this
excess is generated by pulsars, since the flux from the dark matter is
expected to be smooth with a single cutoff.  If these features are not
observed, a pulsar explanation of the observed excess is possible if
there are no young energetic pulsars in the vicinity of the Earth with
\hbox{$E_{\rm br} \ll E_{\rm inj}$} \cite{Grasso:2009ma} or there are
considerable spatial variations in the energy losses of these particles
as they diffuse through the ISM.  The amplitude of these fluctuations
also depends on the relative contributions of the backgrounds and the
pulsars.  If backgrounds dominate at high energies, these fluctuations
may be undetectable.


This paper is organized as follows: In Sec. \ref{sec:single}, we
review the propagation of electrons in the ISM.  We estimate the
typical propagation time and distance for electrons and positrons of a
given energy and show that injection of electrons and positrons by a
pulsar can be approximated by delta-functions in space and time.
Using the properties of pulsars and their PWN derived in
Appendix \ref{Pulsars-app}, we derive the temporal evolution of the
observed flux from a single pulsar at a given distance.  In
Sec. \ref{sec:distr}, we study the flux from a distribution of
pulsars, calculating the average expected flux and comparing this
result to the estimate flux from pulsars in the Australia Telescope National Facility
(ATNF) catalog
\cite{Manchester:2004bp} -- demonstrating that at energies $E \lesssim
300$ GeV this flux is well approximated by the average curve while at
higher energies there are significant deviations and bumplike
features due to the cooling breaks.  We then compare both spectra with
the ATIC, {\it Fermi}-LAT, and {\it PAMELA} data.  In
Sec. \ref{sec:DM}, we review the flux from dark matter.  In
Sec. \ref{sec:conclusion}, we present our conclusions.  We argue that
the flux from dark matter is likely indistinguishable from the flux
from a single pulsar or from a continuous distribution of pulsars.
Thus, unless there are additional features at high energies that point
to the contribution from several pulsars, it may be impossible to tell
whether the excess is due to dark matter or pulsars.

The paper also contains a few appendixes:
In Appendix \ref{Pulsars-app} we give a general review of pulsars and
pulsar wind nebulae.  In Appendix \ref{Smearing-app} we derive the
smearing of the cooling breaks due to spatial variability of energy
losses.  In Appendix \ref{constr-app} we study some constraints that
current electron and positron data put on the pulsar models.


\section{Single pulsar flux}
\lb{sec:single}

The flux of cosmic ray electrons at the Earth depends on both the
spectrum of injected electrons and the properties of the ISM.  First,
we review the propagation of electrons in the ISM, and then derive the
expected flux from pulsars and dark matter.

\subsection{Properties of the interstellar medium}
\lb{sec:prop}

The ISM contains a magnetic field with a strength on the order of
$3\mu \text{G}$ \cite{Longair1992}.  Since the corresponding Larmor
radius for a 1 TeV electron is small, $r_L = \frac{pc}{eB} < 10^{-3}$
pc, electrons are expected to mostly follow the ISM's magnetic field
lines.  Because the ISM magnetic field has random fluctuations,
electrons propagate along on a random path.  The
corresponding diffusion coefficient for relativistic particles is
\cite{Strong:2007nh} 
\be 
D (E) = D_0\left(\frac{E}{E_0}\right)^\dl,
\ee 
where $\dl = 0.3 - 0.6$ and $D_0 = (3 - 5) \times 10^{28}
\text{cm}^2\: \text{s}^{-1}$ for $E_0 = 1$ GeV.  [The typical mean
free path for a 1 TeV electron $r_{\rm f} \sim D(E) / c > 1$ pc is
much larger than the Larmor radius $r_{\rm L} < 10^{-3}$ pc.] We find
it convenient to express the diffusion coefficient in terms of the
energy rather than the magnetic rigidity, $R \equiv p/q$, where $p$ is
the momentum and $q$ is the charge of the particle.  In our case, the
two definitions are equivalent.

As electrons propagate in the ISM, they lose energy.  For electrons
with energy $E \gtrsim 5$ GeV, the dominant loss mechanisms are
synchrotron radiation and inverse Compton scattering off cosmic
microwave background, infrared (IR), and starlight photons.  The
corresponding energy losses are 
\be \lb{Eloss} 
\dot E \equiv - b (E) =
- b_0 E^2, 
\ee 
where $b_0 = 1.6\times 10^{-16} \text{GeV}^{-1}
\text{s}^{-1}$ for the local density of photons \cite{Porter:2005qx}
and $B = 3\: \mu \text{G}$.  Before we present a formal solution to
the propagation equations, we define a few characteristic numbers.  In
estimations, it is convenient to represent the parameters in the units
of pc $= 3 \times 10^{18} $cm and kyr = $3 \times 10^{10} $s.  In this
paper we will usually use $b_0 = 1.6\times 10^{-16} \text{GeV}^{-1}
\text{s}^{-1} = 5\times 10^{-6} \text{GeV}^{-1} \text{kyr}^{-1}$ and
$D_0 = 3 \times 10^{28}\text{cm}^2 \text{s}^{-1} = 100\, \text{pc}^2\,
\text{kyr}^{-1}$ with $\dl = 0.4$.
%

By integrating Eq.~(\ref{Eloss}), we find the energy loss in terms of
the electron travel time is 
\be \lb{coolE} 
\frac{1}{E_1} - \frac{1}{E_0} = b_0 t, 
\ee 
where $E_0$ is the initial energy of the
electron and $E_1$ is the energy at time $t$.  The {\it cooling
break}, defined as the maximal energy an electron can have after
traveling for time $t$, is
\be \lb{cbr} 
E_{\rm br} = \frac{1}{b_0 t}.  
\ee 
The characteristic travel time is therefore 
\be \lb{ctime} 
t \gtrsim 100\:\text{kyr} \qquad \text{for} 
\qquad E \lesssim 2\: \text{TeV}.
\ee 
The characteristic distance an electron travels before cooling to
energy $E$ is the diffusion distance $x_{\rm diff}^2 = 4 D(E) t$,
where $t = \frac{1}{b_0 E}$, 
\be \lb{cdist} 
x_{\rm diff} \lesssim 5\,\text{kpc} 
\qquad \text{for} \qquad E \gtrsim 10\, \text{GeV}.  
\ee

\subsection{Green function for diffusion-loss propagation}

In general, the evolution of the energy density $\rho$
of electrons moving in random paths and losing energy can be described
by the following diffusion-loss equation
\cite{Longair1992}\cite{Ginzburg1964}
\be 
\lb{diff-loss} 
\frac{\p\rho}{\p t} = \frac{\p }{\p E} \text{\Large (}b (E) \rho \text{\Large )}
+ \frac{\p}{\p x^i} \text{\Large (}D (E) \frac{\p}{\p x^i} \rho \text{\Large )} 
+ Q(\vx, E, t), 
\ee 
where $Q \equiv dN/(dE dt d^3x)$ is the energy density of electrons
injected by the source.  In principle, one can also take into account
reacceleration, convection, and decays (collisions), but for
electrons with $E > 10$ GeV these contributions can be ignored.

The general solution to Eq.~(\ref{diff-loss}) is found in
\cite{Ginzburg1964}\cite{1959AZh....36...17S}.  To solve
Eq.~(\ref{diff-loss}) for a general source, one introduces the Green
function $G (\vx,\: E,\: t;\: \vx_0,\: E_0,\: t_0)$ which satisfies
\bea
\nonumber
&&\frac{\p G}{\p t} 
- \frac{\p }{\p E} \text{\Large (}b (E) G \text{\Large )} 
- D(E) \frac{\p^2 G}{\p x^2}  =  \\
\lb{Green}
&&\dl(\vx - \vx_0) \dl(E - E_0)\dl(t - t_0).  
\eea 
Then, the solution to (\ref{diff-loss}) is 
\bea 
&& \nonumber
\rho (\vx,\: E,\: t) = \\ 
&& \nonumber
\int d^3 \vx_0 \int dE_0 \int dt_0\; G (\vx,\: E,\: t;\:
\vx_0,\: E_0,\: t_0)\\ 
&&\lb{densol}
\cdot Q (\vx_0,\: E_0,\: t_0).  
\eea
The Green function can be derived as follows.  One can define the
variables $t' = t - \tau$ and $\ld$
\cite{Ginzburg1964}\cite{1959AZh....36...17S}, where 
\bea
\lb{new-vars1}
\tau \equiv \tau(E,\: E_0) = \int_E^{E_0} \frac{dE'}{b (E')},\\
\lb{new-vars2}
\ld   \equiv \ld(E,\: E_0)   = \int_E^{E_0} \frac{D (E')dE'}{b (E')}.
\eea
The variable $t'(t, E)$ is invariant with
respect to the differential operator $\p_t - b (E) \p_E$.  In fact,
$D^{-1}(E)(\p_t - b (E) \p_E) = \p_\ld$ and Eq. (\ref {Green})
becomes the usual diffusion equation in $\ld$ and $\vx$.
The Green function is then \cite{Ginzburg1964}\cite{1959AZh....36...17S}
\bea
\lb{GeneralGreen}
&&G (\vx,\: E,\: t;\: \vx_0,\: E_0,\: t_0) 
= \frac{1}{b (E)} \frac{1}{(4\pi\ld)^{3/2}} e^{-\frac{(\vx - \vx_0)^2}{4\ld}}\nonumber\\
&&\cdot \dl(t - t_0 - \tau)
\theta(E_0 - E).
\eea
Equation~(\ref{diff-loss}) and the above Green function have a few
limitations.  Both the ISM magnetic field and density of IR and
starlight photons vary in space. Consequently the diffusion
coefficient and the energy loss function depend on the coordinates: 
$D = D(E,\: \vx)$, $b = b(E,\: \vx)$, and there is no simple analytic
solution to Eq.~(\ref{diff-loss}).  In Appendix~\apref{Smearing-app} we
calculate corrections to the predicted $e^+e^-$ spectrum at the Earth
due to spatial variations in the energy loss function.

\subsection{Flux from a single pulsar}
\lb{sec-single}

With the general Green function in hand, one can find the density of
electrons at any point in space for any source.  In this Section, we
derive the expected flux of electrons and positrons produced by a
single pulsar.

The distances to pulsars are sufficiently large that we can assume
that pulsars are point sources.  We also assume that most of the
pulsars' rotational energy is lost via magnetic dipole radiation
\cite{Shapiro1983} which eventually transforms into the energy of
electrons and positrons 
\be
\lb{pulsar-source}
Q_{\rm pulsar}(\vx, E, t) = Q(E)\,
\frac{1}{\tau}\left(1 + \frac{t}{\tau}\right)^{-2}\; \theta(t) \;
\dl(\vx),
\ee
where $t$ is the pulsar age and $\vx$ is its position.
$\theta(t)$ is the step function that ensures $Q_{\rm pulsar} = 0$ for $t < 0$.
Note that the pulsar spin-down time scale $\tau$ in
this formula and the variable introduced in (\ref{new-vars1}) are
unrelated.  We review the derivation of this formula in
Appendix~\apref{Pulsars-app}.

At late times ($t\gg\tau$), the spin-down luminosity scales as
$t^{-2}$.  Consequently, most of the energy is emitted during $t \sim
\tau$.  The pulsar spin-down time scale $\tau \lesssim 10$ kyr is
much smaller than the typical electron propagation time $t \gtrsim
100$ kyr.  Consequently, we can take the limit $\tau \ra 0$, which
results in 
\be
\lb{delta-fun-appr}
\left. \frac{1}{\tau}\left(1 + \frac{t}{\tau}\right)^{-2}\theta(t) \right|_{\tau \ra 0} 
\lra \;\; \dl(t).
\ee
For pulsars with a significant PWN, the time dependence in
Eq. (\ref{pulsar-source}) does not describe the escape of electron and
positrons from the PWN into the ISM.  However, for most pulsars the
lifetime of the PWN ($t \lesssim 20$ kyr; \cite{Gaensler:2006ua}) is
much smaller than the typical propagation time $t \gtrsim 100$ kyr
[Eq. (\ref{ctime})].  Thus, the detailed time dependence of the escape
is not significant and the delta-function approximation in
Eq. (\ref{delta-fun-appr}) is still valid.

We assume that the energy spectrum of particles injected into the ISM
$Q(E)$ by a given pulsar has the form \be Q(E) = Q_0\: E^{-n}\: e^{-
\frac{E}{M}}, \ee where $n$ is the injection index and $M$ is the
injection cutoff.  We denote the initial rotational energy of the
pulsar by $W_0$ and define $\eta$ to be the fraction of this energy
deposited in the ISM as $e^+e^-$.  The total energy emitted in
$e^+e^-$ is then 
\be 
\int Q(E) E dE = \eta W_0.  
\ee 
For $n < 2$, this gives 
\be \lb{normDef} 
Q_0 = \frac{\eta W_0}{\G(2 - n) M^{2 - n}}.
\ee 
The index $n$ and the cutoff $M$ of the electron spectrum can be
derived from the broadband spectrum of a PWN, and may vary
significantly between pulsars.  We argue in Appendix~\apref{Pulsars-app}
that reasonable values for energetic pulsars are $n = 1.5\pm0.5$ and
$M \sim 100\:\text{GeV} - 10\:\text{TeV}$.

The overall normalization is more difficult to derive because neither
the initial rotational energy nor the conversion efficiency are known
for most pulsars.  Currently, models can only estimate $W_0$ to an
order of magnitude with significant theoretical uncertainties
\cite{FaucherGiguere:2005ny}.  In Appendix~\apref{Pulsars-app}, we derive
that, assuming a constant pulsar time scale $\tau = 1$ kyr for all
pulsars in the ATNF catalog \cite{Manchester:2004bp}, the distribution
of pulsar initial rotational energies $W_0 \equiv 10^p\:$erg satisfies
a log-normal distribution with $\bar p \approx 49$ and $\sm_p \approx
1$ which gives the average $\bar W_0 \approx 10^{50}\:$erg.  If we use
$\tau = 10$ kyr, the same analysis gives $\bar p \approx 48$, $\sm_p
\approx 1$, and $\bar W_0 \approx 10^{49}\:$erg.  If the age of a
pulsar is known independently, then the initial rotational energy can
be estimated more robustly.  For example, the Crab pulsar is
associated with the SN1054 supernova explosion and has $\tau \approx
0.7$ kyr and $W_0 \approx 5.3\times 10^{49}$ erg.

Let us now discuss the conversion coefficient $\eta$.  The energy
density near the surface of the pulsar is dominated by the magnetic
field and the spin-down luminosity is dominated by the magnetic dipole
radiation.  In most PWNe the energy density is believed to be particle
dominated, i.e.  at large distances from the neutron star most of the
energy outflow has been converted to particles and $\eta \sim 1$ at
this stage (see, e.g., \cite{1996MNRAS.278..525A} for a discussion of the
Crab PWN).  However, these particles do not immediately escape to the
ISM but are trapped inside the PWN by its magnetic field where they
can lose a significant fraction of their energy.  In
Appendix~\apref{Pulsars-app}, we estimate $\eta \sim 0.1$ due to cooling
of the particles before they escape into the ISM.  Based on the
discussion above, we find that the average energy in electrons and
positrons $\eta W_0 \sim 10^{49}$~erg is reasonable.  This value is
model dependent and can vary greatly from one pulsar to another.

The density of electrons propagated from a pulsar to the Earth can be
found by substituting the source function $Q (\vx,\: E,\: t)$ into
Eq.~(\ref{densol}) 
\be
\lb{psr-dens}
\rho(\vx,\: E,\: t)
= \frac{b(E_0)}{b(E)}
\frac{1}{(4\pi\ld)^{3/2}} e^{-\frac{\vx^2}{4\ld}} 
Q(E_0),
\ee
where parameter $\ld$ is defined in Eq.~(\ref{new-vars2})
and $E_0$ is the initial energy of the electrons that cool down to $E$
in time $t$ 
\be 
E_0 = \frac{E}{1 - E b_0 t}.  
\ee 
The density in Eq. (\ref{psr-dens}) has a cutoff at the cooling break, 
$E = \frac{1}{b_0t}$, since $Q(E_0) \ra 0$ for $E_0 \ra \infty$.

\begin{figure}[tp] 
\begin{center}
\epsfig{figure = 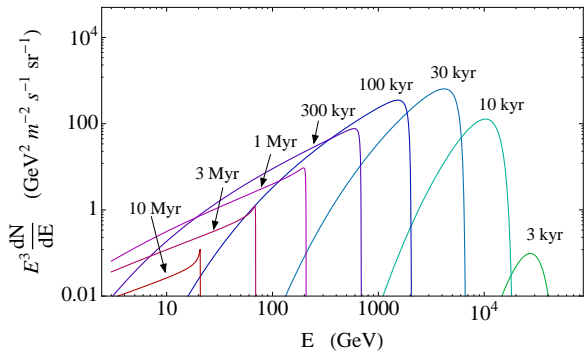,scale=.4}
\vspace{-2mm}
\end{center}
\noindent
\caption{\small Time evolution of $e^+e^-$ flux on the Earth from a
pulsar at a distance of 1 kpc with $\eta W_0 = 3\times 10^{49}\:$erg,
an injection index $n = 1.6$, and an injection cutoff $M = 10$ TeV.
The diffusion and energy losses are described in Sec. \ref{sec:prop}.
We assume the delta-function approximation for the emission from the
pulsar, $Q(\vx, E, t) = Q(E) \dl(\vx)\dl(t)$.  The flux from a young
pulsar (the 3 kyr curve on the right) has an exponential suppression
because the electrons have not had enough time to diffuse from the
pulsar to the Earth.  The cutoff moves to the left due to cooling of
electrons and becomes sharper.  After reaching a maximal value, the
flux decreases since the electrons diffuse over a large volume.  }
\label{DensityEvolution}
\vspace{-2mm}
\end{figure}

For a density $\rho$ of relativistic particles, the flux is defined as
\be 
F = \frac{c}{4\pi}\rho.  
\ee 
The time evolution of the flux from a
single pulsar is shown in Fig.~\ref{DensityEvolution}.  At early
times, the electrons have not had enough time to diffuse to the
observer and the flux is exponentially suppressed.  At later times,
the flux grows until the diffusion distance $\sqrt{4 D(E) t}$ is
similar to the distance from the pulsar to the Earth.  After that the
flux decreases as the electrons diffuse over a larger volume.  The
cutoff moves to lower energies due to cooling of electrons.

\begin{figure}[tp]
\begin{center}
\epsfig{figure = 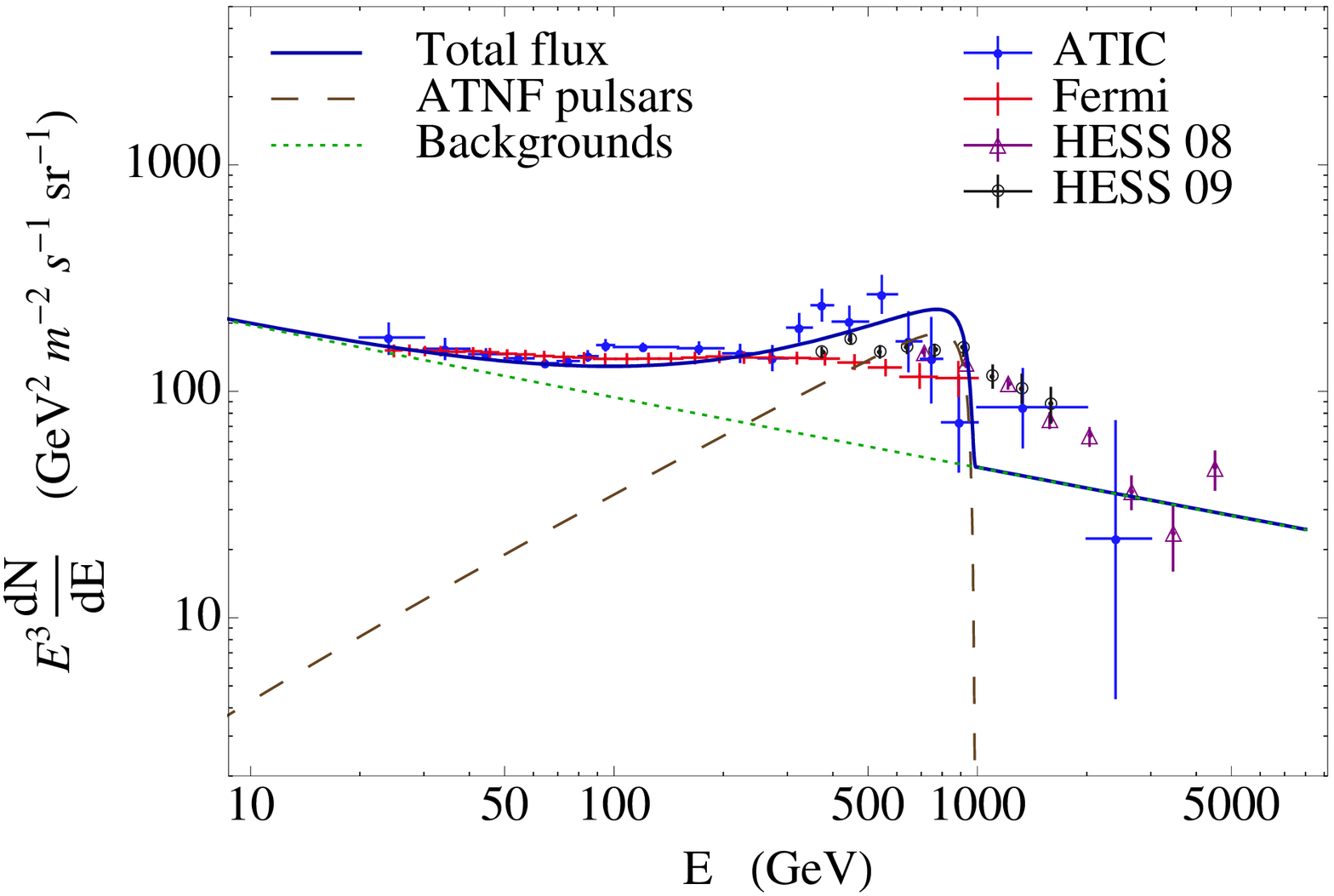,scale=0.4}
\hspace{3mm}
\epsfig{figure = 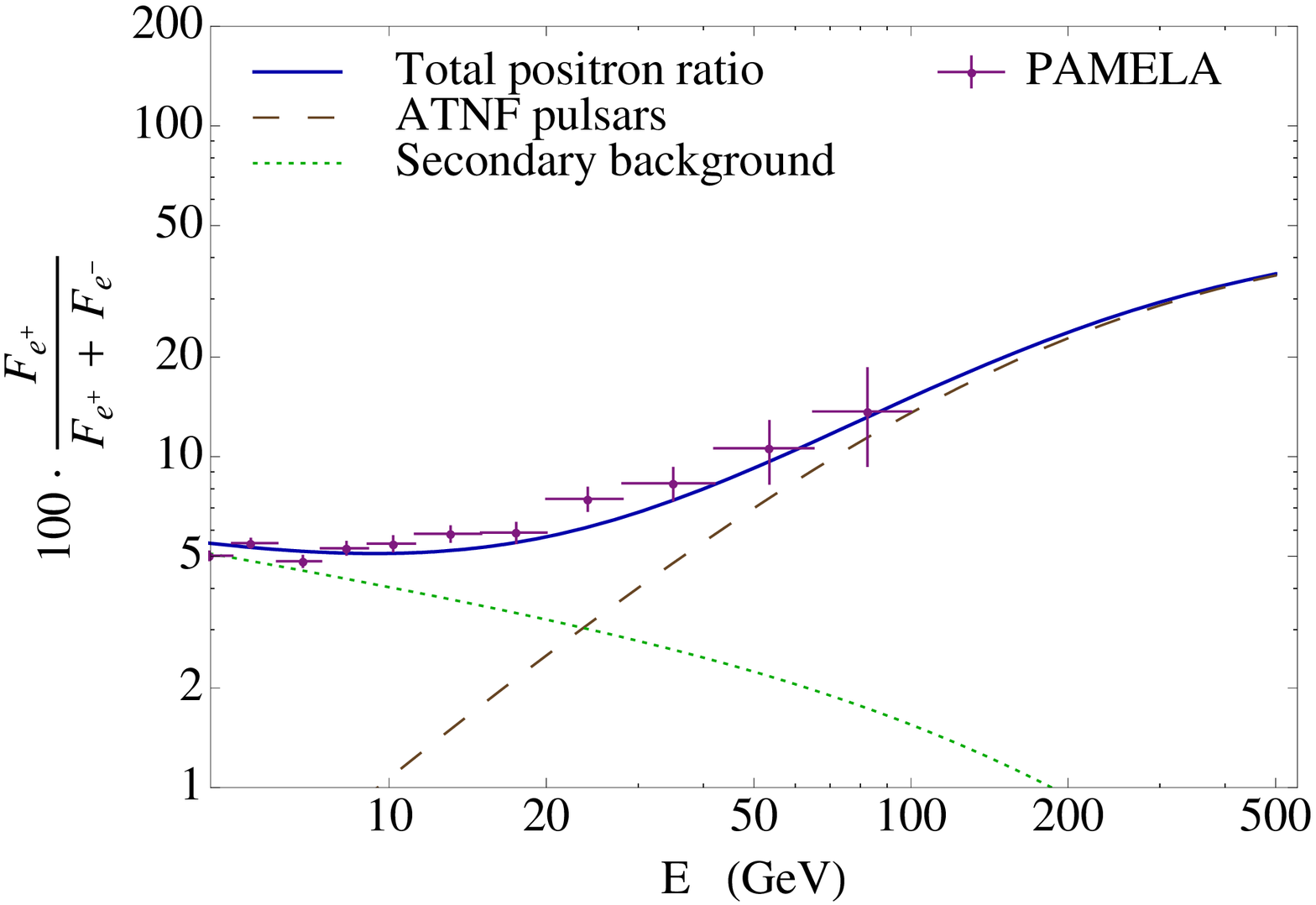,scale=0.4}
\vspace{-2mm}
\end{center}
\noindent
\caption{\small Electron and positron flux from a single pulsar
together with a primary background $\sim E^{-3.3}$ and a secondary
background $\sim E^{-3.6}$.  The pulsar is at a distance of 0.3 kpc.
It has $\eta W_0 = 2.2 \times 10^{49}$ erg, age of 200 kyr, and injection
index and cutoff $n = 1.6$ and $M = 10$ TeV, respectively.  The propagation
parameters are described in Sec. \ref{sec:prop}.  The cutoff $M \gg 1$
TeV results in a significant bump around 1 TeV which is consistent
with the ATIC data.  For a smaller injection cutoff $M \sim 1$ TeV,
the flux from the pulsar takes the form of a power law with an
exponential cutoff that can be used to fit the {\it Fermi} and {\it
PAMELA} data (see, e.g., \cite{Malyshev:2009zh}).  }
\label{SinglePsr}
\end{figure}

For energies much smaller than the cooling break, we can neglect
energy losses.  In this case, $E_0 \approx E$, $\ld \approx D(E) t$
and Eq. (\ref{psr-dens}) reduces to 
\be \lb{psr-dens-diff}
\rho(\vx,\: E,\: t) = \frac{1}{(4\pi D(E) t)^{3/2}}
e^{-\frac{\vx^2}{4D(E) t}}\:Q(E).  
\ee 
Assuming that $\vx^2 \ll 4 D (E) t$, the flux for $E \ll \frac{1}{b_0t}$ is 
\be \lb{DeltaFlux} 
F(E) = \frac{c}{4\pi} \frac{Q_0}{(4\pi D_0 t)^{3/2}} E^{- n - \frac{3}{2}\dl}.  
\ee 
In general, 
the flux that we add to the backgrounds to fit
the data can be effectively parametrized by three numbers: the
normalization, the index at low energies, and the cutoff energy.  From
the right hand side of Eqs. (\ref{normDef}) and (\ref{DeltaFlux}) we
find that these three parameters correspond to at least 8 parameters
describing the pulsar and the ISM.  In particular, the propagated
index $n_a$ is a linear combination of $n$ and $\dl$, $n_a = n +
\frac{3}{2} \dl$, the propagated cutoff is $E_{\rm cut} = \frac{1}{b_0
t}$, and the normalization depends on $\eta$, $W_0$, $M$, $D_0$, and
$t$.  In order to fix this degeneracy, as a matter of convenience, we
choose $D_0 = 3\times 10^{28}\text{cm}^2\text{s}^{-1}$, $\dl = 0.4$,
$b_0 = 1.6\times 10^{-16} \text{GeV}^{-1} \text{s}^{-1}$, $W_0 =
10^{50}\:$erg, and $M = 10$ TeV.  With this choice, our fit to the
$e^+e^-$ data will determine $n$, $\eta$, and $t$.  If some of the
parameters are known independently, e.g., the propagation model, the
energy losses, the age of the pulsar etc., this approach becomes more
constrained and more predictive.  As shown in Fig. \ref{SinglePsr},
the expected flux from a pulsar with $\eta W_0 \approx 3\times
10^{49}$~erg, $n = 1.6$, distance 0.3 kpc, and age 200 kyr reproduces
the positron fraction measured by {\it PAMELA} and is a good fit to the
cosmic-ray electron spectrum measured by ATIC, {\it Fermi}, and HESS
below $\sim1$ TeV.  This suggests that the anomaly in the $e^+e^-$
flux could be due to a single pulsar.  However, given the considerable
number of known nearby, energetic pulsars \cite{Manchester:2004bp}, it
is unlikely that the flux from any single pulsar is significantly
larger than the flux from all such pulsars.  In the next section, we
will derive the expected flux of electrons and positrons from a
collection of pulsars.


\section{Flux from a collection of pulsars}
\lb{sec:distr}

In this section, we derive the $e^+e^-$ flux from a continuous
distribution of pulsars and compare it with the predicted flux from
the pulsars in the ATNF catalog \cite{Manchester:2004bp}.

\subsection{Flux derivation}

We assume that pulsars are homogeneously distributed
in the galactic plane and are born at a constant rate $N_b$
\cite{FaucherGiguere:2005ny}. 
The ``continuous'' distribution of pulsars is defined 
as the average of all possible realizations of
pulsar distributions.  
This results in a source function constant in
time, localized in the vertical direction, and homogeneous in the
galactic plane 
\be \lb{PulsarDistr} 
Q_{\rm distr}(\vx, E, t) = J_0\: E^{-n}\: e^{- \frac{E}{M}}\, \dl(z) 
\ee 
with the normalization constant 
\be 
J_0 = \frac{\eta W_0}{\G(2 - n) M^{2 - n}} \frac {N_b} {A_{\rm gal}}, 
\ee 
where $A_{\rm gal}$ is the area of the galactic
plane.  Since the diffusion distance of these electrons is
significantly smaller than the distance from the Earth to the edge of
the galactic plane \cite{FaucherGiguere:2005ny} ($x_{\rm diff} < 10$
kpc), we can neglect the effects of having an edge at a finite
distance.

\begin{figure}[t] 
\begin{center}
\epsfig{figure = 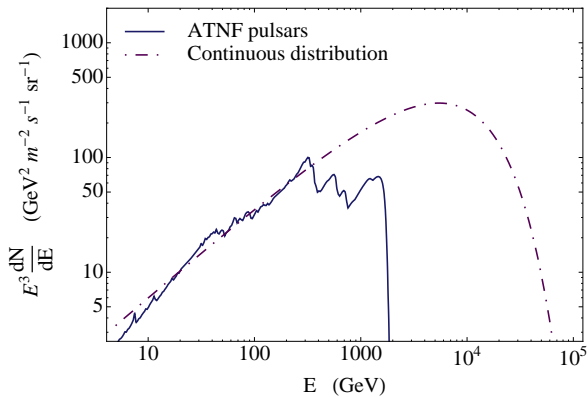,scale=.4}
\vspace{-2mm}
\end{center}
\noindent
\caption{\small The expected spectrum from the continuous flux
  distribution and that from pulsars in the ATNF catalog pulsar
  \cite{Manchester:2004bp}.  The latter is calculated using $\eta =
  0.065$, $n = 1.5$, and a pulsar time scale $\tau = 1$ kyr for each
  pulsar. This last fact, in conjunction with its spin-down age and
  current spin-down luminosity, is used to calculate each pulsar's
  initial rotational energy through Eq.~(\ref{iniW0}).  We also use
  the value of the propagation parameters given in
  Sec.~\ref{sec:prop}.  Several hundred pulsars contribute below 300
  GeV and the continuous distribution provides a good approximation
  for these energies.  Above 300 GeV, there are only $\sim10$
  contributing pulsars, and the observed flux in this energy range is
  strongly dependent on their individual properties.  The reason for
  the significant discrepancy between these two curves above 2 TeV has
  to do with the actual local distribution of pulsars versus the
  averaged flux seen by many observers in the Galaxy, as discussed in
  Sec.~\ref{sec:stat_cut}.} 
\label{4ATNFdistr}
\end{figure}

\begin{figure}[tp]
\begin{center}
\epsfig{figure = 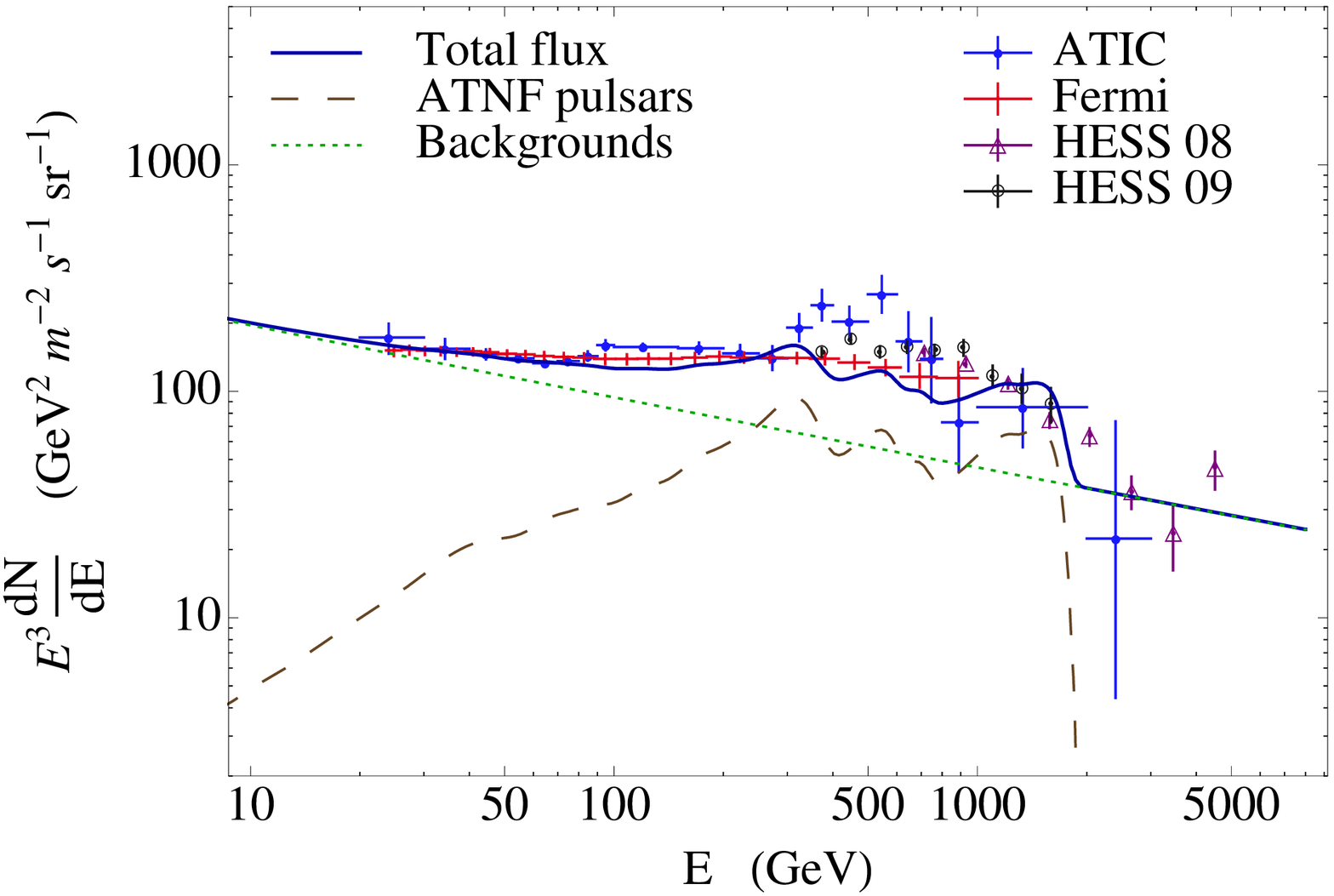,scale=0.4}
\hspace{3mm}
\epsfig{figure = 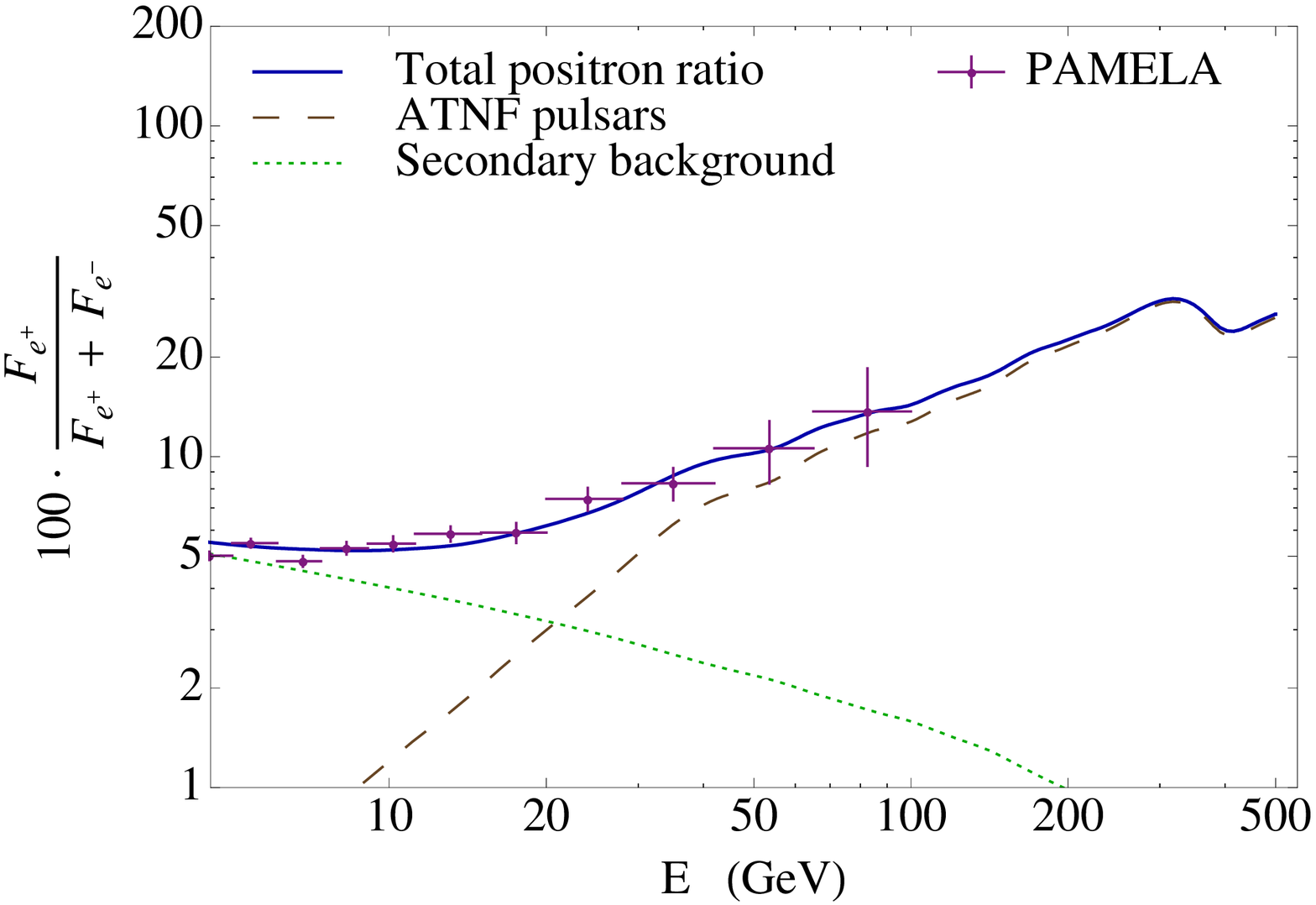,scale=0.4}
\vspace{-2mm}
\end{center}
\noindent
\caption{\small The predicted flux from pulsars in the ATNF catalog
  calculated using the same procedure as in
  Fig. \ref{4ATNFdistr} but accounting for spatial variations of
  energy losses as described in Appendix \ref{Smearing-app}.  The assumed
  backgrounds are the same as in Fig. \ref{SinglePsr}.}
\label{PsrsFig}
\end{figure}

Using the general Green function in Eq. (\ref {GeneralGreen}),
the flux of electrons from this distribution is 
\bea
&& F = \nonumber\\
&&\frac{c}{4\pi}
\int d^3 \vx_0 \int dE_0 \int dt_0\; G (\vx,\: E,\: t;\: \vx_0,\: E_0,\: t_0)\nonumber\\
&&\cdot Q (\vx_0,\: E_0,\: t_0).
\eea
Integrating over $t_0$ and $\vx$, we obtain 
\be
\lb{DistributionFlux}
F (E) = \frac{c}{4\pi b (E)}
\int_E^\infty dE_0 \frac{1}{\sqrt{4 \pi \ld(E,\: E_0)}} J_0 E_0^{-n}\: e^{- \frac{E_0}{M}},
\ee
where $\ld$ is defined in Eq.~(\ref{new-vars2}).
This flux can be rewritten as 
\be
\lb{ContDistrScaling}
F_{distr}(E) = \frac{c}{4\pi} \frac{J_0}{\sqrt{4\pi b_0 D_0}}\; 
I_\frac{E}{M}\;
E^{- n - (\dl + 1) / 2},
\ee 
where
\be
\lb{norm-const}
I_\frac{E}{M} = \int_1^{\infty} d x \sqrt{\frac{1 - \dl}{1 - x^{\dl - 1}}} \; x^{-n} \: 
e^{- \frac{E}{M}x},
\ee
for example, if $E \ll M$, $\dl = 0.4$, and $n = 1.5$, 
then $I_\frac{E}{M} \approx 3$.

As in the case of a single pulsar flux, the number of parameters we
need to fit the data is much smaller than the number of parameters
characterizing the flux from a collection of pulsars.  In this case,
the index of the observed flux and the normalization can be found from
Eq. (\ref{ContDistrScaling}).  For example, the index of the flux at
low energies $n_a = n + (1 + \dl)/2$.  
Formally, the cutoff in this case is equal to the injection
cutoff $M$, but for an actual distribution of pulsars the expected
cutoff is lower and is determined by the age of the youngest pulsar
within the diffusion distance from the observer -- as derived in
Sec.~\ref{sec:stat_cut}.  If we break the degeneracy by picking a
particular propagation model, we can constrain the properties of the
pulsar distribution.  The opposite is also true: by choosing some
properties of the pulsars one can constrain the properties of the ISM
-- as demonstrated in Appendix \ref{constr-app}.

In order to break the degeneracy we fix the ISM properties as in
Sec. \ref{sec-single}.  To calculate the flux from the pulsars in
the ATNF catalog we use the following toy model.  We assume that every
pulsar has injection index $n = 1.5$ and conversion efficiency $\eta =
0.065$ (these values are chosen to fit the low energy electron and
positron data in Fig. \ref{PsrsFig}).  We choose an injection cutoff
$M = 10\:$TeV for every pulsar (for smaller values of $M$ the features
at high energies will be less sharp, since the injection cutoff is not
as abrupt as the cooling break).  In order to estimate the initial
rotational energy, we assume that for each pulsar, the spin-down time
scale is $\tau$ = 1 kyr.  Then we use Eq.~(\ref{EdotEvolution}) to
express the initial rotational energy $\mcE_0 \equiv W_0$ in terms
of the current spin-down luminosity $\dot\mcE$ and the pulsar age $t
\gg \tau$:
\be 
\lb{iniW0} W_0 \approx \dot\mcE \frac{t^2}{ \tau}.  
\ee
The result in shown in Fig. \ref{4ATNFdistr}, and the relative
normalization between this spectrum and that of the continuous
distribution described above depends on the pulsar birth rate $N_{\rm
b}$, or, to be more precise, on the local value of the pulsar birth
rate.  In order to have a good agreement between the two curves for
energies 30 -- 300 GeV, we require $N_{\rm b} \approx 1.8\:
\text{kyr}^{-1}$, assuming a Milky Way radius $R_{gal}$ = 20 kpc
\cite{FaucherGiguere:2005ny}.  For energies below 30 GeV, we find that
the main contribution to $e^+e^-$ flux comes from the pulsars with age
$t > 10$ Myr.  These pulsars typically have a very low spin-down
luminosity and therefore are difficult to observe (in the ATNF catalog
there are very few pulsars with the spin-down luminosities $\dot E <
10^{31}$ erg).  In Fig. \ref{PsrsFig}, we apply to this spectrum the
Gaussian smearing expected to result from spatial variations in energy
losses depending on the path of the electrons -- as derived in
Appendix \ref{Smearing-app}.  As one can see, this provides a very good
fit to the {\it PAMELA}, {\it Fermi}, and HESS data - but does not
reproduce the ATIC bump.

Determining the flux of $e^+e^-$ from the actual distribution of
pulsars using a more realistic model is extremely difficult because
every pulsar has its own independent parameters (e.g., $W_0$, $\eta$, and
$\tau$).  Thus, we may choose several thousands of parameters in order
to fit less than a hundred of data points (which can be fitted by a
flux parametrized by three parameters only).  Moreover, as we
discussed in the Introduction, these thousands of parameters refer to
PWNe sufficiently old that their electrons have had enough time to
diffuse to the Earth.  These PWNe have already disappeared and
therefore cannot be observed directly -- making it impossible to
directly constrain these parameters observationally.  The large number
of pulsars and the impossibility to derive their individual properties
suggest a statistical method is needed to study the $e^+e^-$ flux
they produce.  At small energies, a lot of pulsars contribute to the
observed flux on Earth and therefore the properties of an individual
pulsar are unimportant.  In this case, the flux should be well
approximated by some average curve -- as demonstrated in
Fig. \ref{4ATNFdistr}.  In this estimate we included all pulsars with
ages $t > 15$ kyr and use the delta-function approximation of source
functions, $Q(\vx,\: E,\: t) = Q(E)\dl(\vx)\dl(t)$
(Sec.~\ref{sec:single}).  The choice of the lower cutoff on the age of
the pulsars is motivated by the fact that young pulsars, such as the
Vela pulsar, usually have a PWN and therefore their electrons have not
escaped yet into the ISM.  At high energies only a few young pulsars
contribute and the deviation from the average curve may be large.  The
presence of features at high energies may serve as a signature of a
collection of pulsars that can distinguish them from a dark matter or
single pulsar origin for these electrons.

\subsection{Statistical cutoff}
\lb{sec:stat_cut}

\begin{figure}[t] 
\begin{center}
\epsfig{figure = 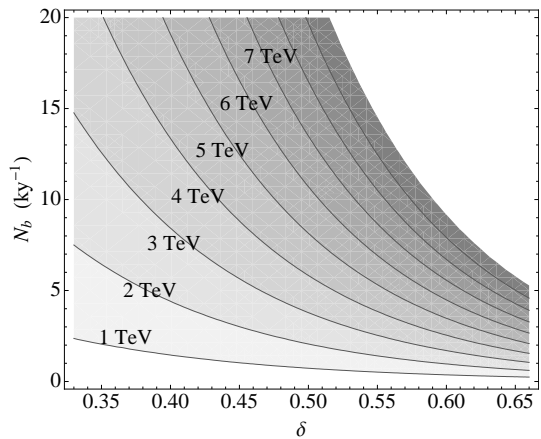,scale=0.4}
\vspace{-2mm}
\end{center}
\noindent
\caption{\small Statistical cutoff as a function of the diffusion
index and the birth rate of pulsars in the Galaxy.  The cutoff in $e^+
e^-$ flux from pulsars is determined by the age of the youngest pulsar
within the diffusion distance from the Earth.  The average such cutoff
is a universal quantity that depends on the properties of ISM (the
energy losses and the diffusion coefficient) and on the pulsar birth
rate, but it is insensitive to the properties of the injection
spectrum from the pulsars.  We assume $D_0 = 100\, \text{pc}^2
\text{kyr}^{-1}$ and $b_0 = 5\times 10^{-6} \text{GeV}^{-1}
\text{kyr}^{-1}$.  }
\label{BrVsDiff}
\end{figure}

\begin{figure}[tp]
\begin{center}
\includegraphics[scale=0.4]{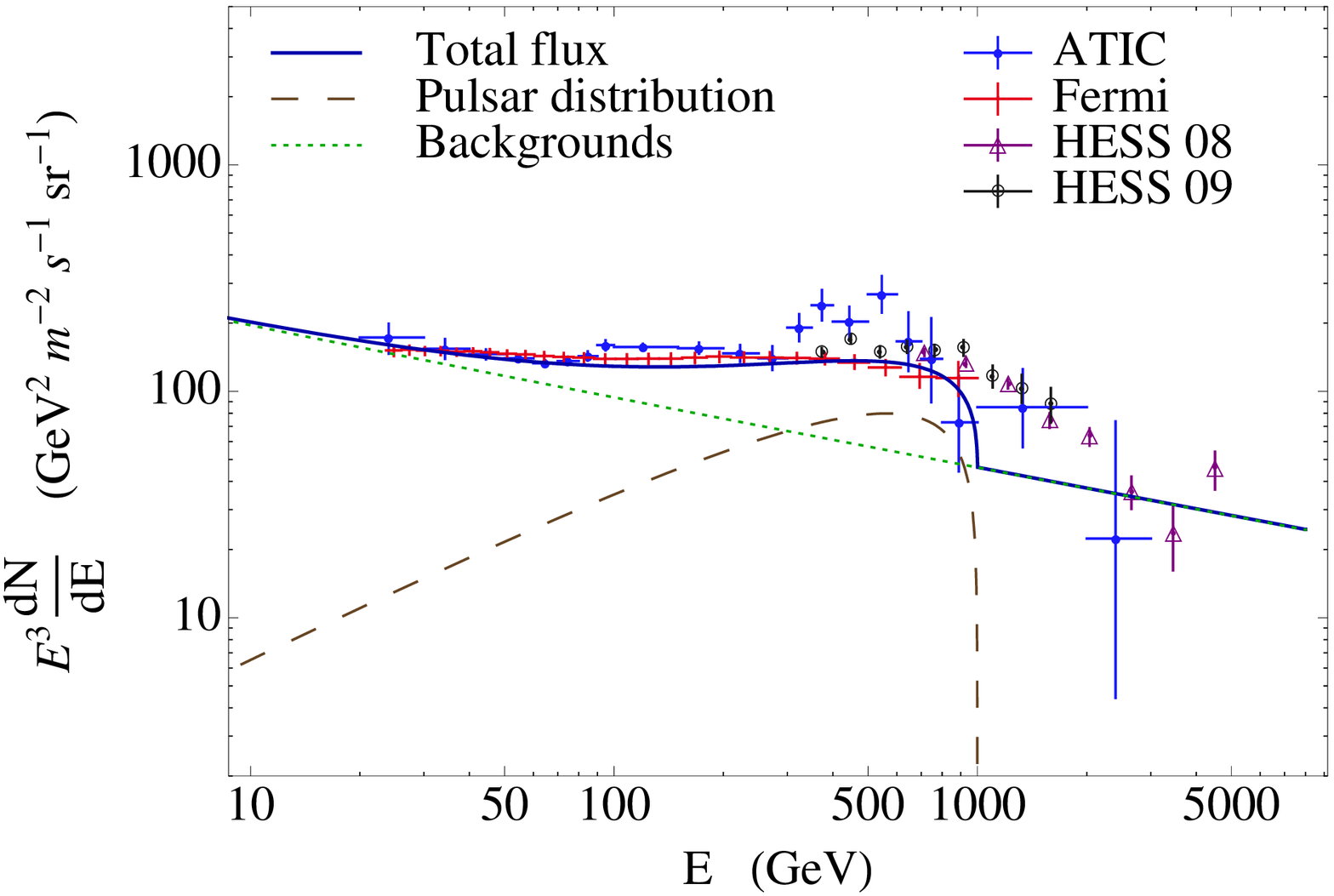}
\hskip 0.15in
\includegraphics[scale=0.4]{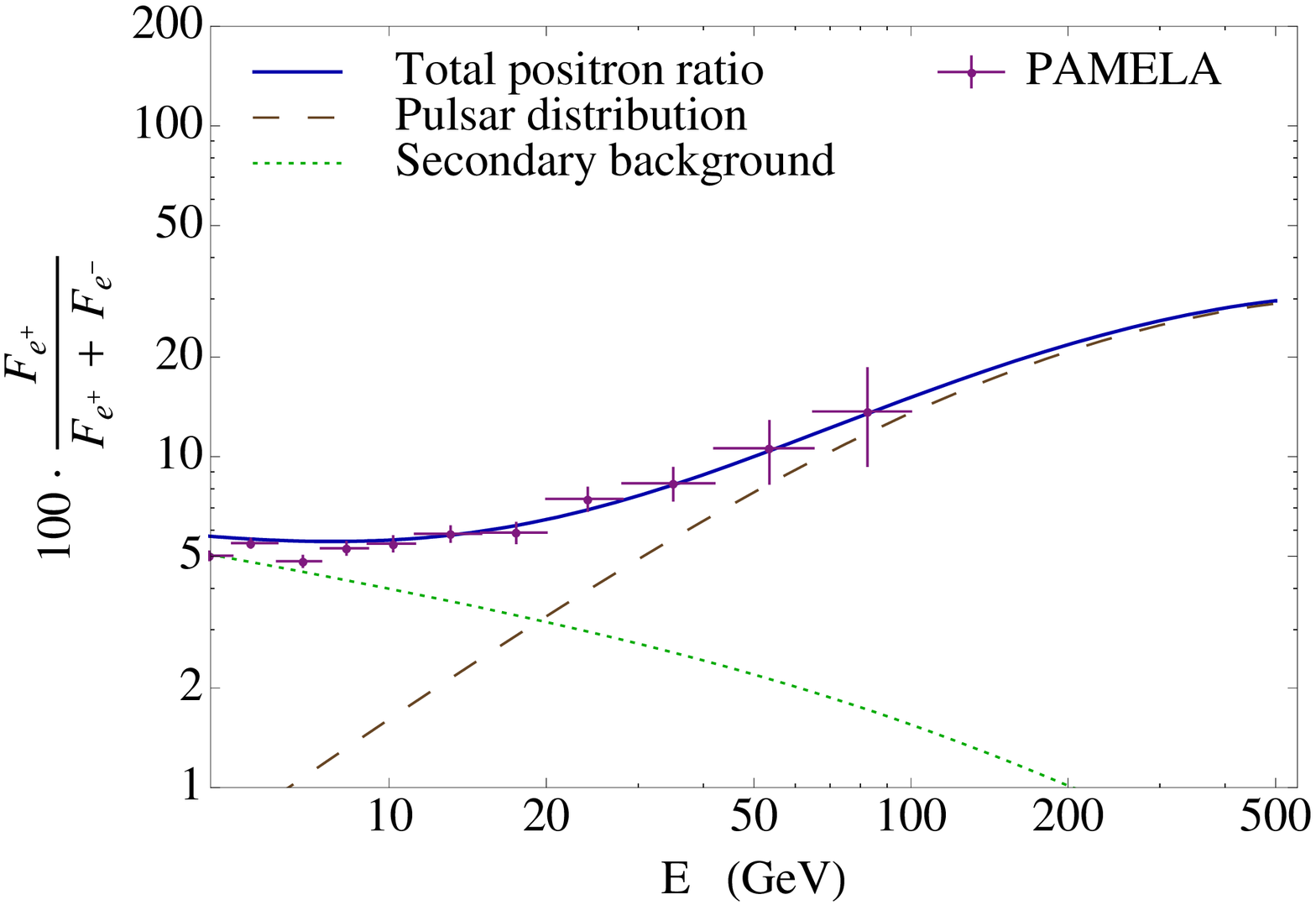}
\vspace{-2mm}
\end{center}
\noindent
\caption{\small The flux from a continuous distribution of pulsars.
The parameters are chosen to fit the {\it Fermi} and {\it PAMELA} data points,
$\eta W_{0}$=$6.5 \times10^{48}$ erg and $n=1.5$.  In this plot,
instead of the injection cutoff $M = 10$ TeV, we use the statistical
cutoff $M_{\rm stat}=1$ TeV.  The backgrounds are the same as in
Fig. \ref{SinglePsr}.  The propagation parameters are described in
Sec. \ref{sec:prop}. 
}
\label{fig:ATIC_PAMELA_fits}
\end{figure}

As shown in Fig.~\ref{4ATNFdistr}, the expected flux from a continuous
distribution of pulsar increases with energy until a break at $E \sim
M = 10$ TeV, whereas the predicted flux from pulsars in the ATNF
database has a cutoff at 2 TeV.  This discrepancy is due to the rare
events when a young pulsar is very close to the observer.  Since
electrons lose energy during propagation, high energy electrons must
come from young pulsars and the cutoff energy is determined by the age
of the youngest pulsar sufficiently close to the observer so that the
electrons have enough time to diffuse through the ISM.  We will call
the average such cutoff a ``statistical'' cutoff.  

To estimate the statistical cutoff, we consider a collection of
pulsars and choose an observation point.  The statistical cutoff at
this point is the maximal cooling break energy for the flux from these
pulsars.  In this distribution, the youngest pulsar whose electrons
can reach the observation point has an age $T$ and diffusion distance
$R$.  For a given pulsar birth rate $N_{\rm b}$, we estimate $M_{\rm
stat}$ by demanding that there is at least one pulsar within $R$
younger than $T$.  Therefore, we have a system of three equations for
the three unknowns $R$, $T$ and $M_{stat}$: 
\bea
M_{\rm stat} = \frac{1}{b_0T},\\
R^2 = 4 D (M_{\rm stat}) T,\\
N_{\rm b}  T\frac{\pi R^2}{A_{\rm gal}} = 1.
\eea
Solving this system of equations, we find
\be
\lb{Mstat-eq}
M_{\rm stat} = \left( \frac{4\pi\:  D_0\: N_{\rm b}}{b_0^2\: A_{\rm gal}}\right)^{\frac{1}{2 - \dl}}.
\ee
Assuming $R_{\rm gal} = 20$ kpc, 
$D_0 = 10^{-4} \text{kpc}^2 \:\text{kyr}^{-1}$,
and $b_0 = 5\times 10^{-6} \text{GeV}^{-1} \text{kyr}^{-1}$, we get
\be
M_{\rm stat} = \left( 4\times 10^5 N_{\rm b} \right)^{\frac{1}{2 - \dl}}\: \text{GeV},
\ee
where $N_{\rm b}$ is in units of $\text{kyr}^{-1}$.
In Fig. \ref {BrVsDiff}, we show the
statistical cutoff as a function of $N_{\rm b}$ and the diffusion
index $\dl$.  This calculation should be viewed as a rough estimate, with
the actual flux from the distribution of real pulsars having a cutoff
that differs by as much as an order of magnitude.  Additionally, it is
possible that current data are missing a feature at high energies
($E\gtrsim2$ TeV) due to poor statistics.
A comparison between the flux from a continuous distribution of pulsars
with $M_{\rm stat} = 1$ TeV and the current data is shown in 
Fig.~\ref{fig:ATIC_PAMELA_fits}.

We note that Eq. (\ref{Mstat-eq}) can also be used to find the cutoff
in the primary background if we assume that it is generated by the
supernova explosions.  For instance, for the supernova rate in the
Milky Way $N_{\rm SN} = 10\:\text{kyr}^{-1}$ and $\dl = 0.4$, it gives
the cutoff in the primary background around 3 TeV.  Using the same
reasoning as above one may expect some features in the spectrum of the
primary electrons at several TeV.  Below $\sim 1$ TeV we do not expect
significant fluctuations in the primary background and the presence of
the features should be interpreted as the signature of pulsars.



\section{Flux from Dark Matter}
\lb{sec:DM}

\begin{figure}[tp]
\begin{center}
\epsfig{figure = 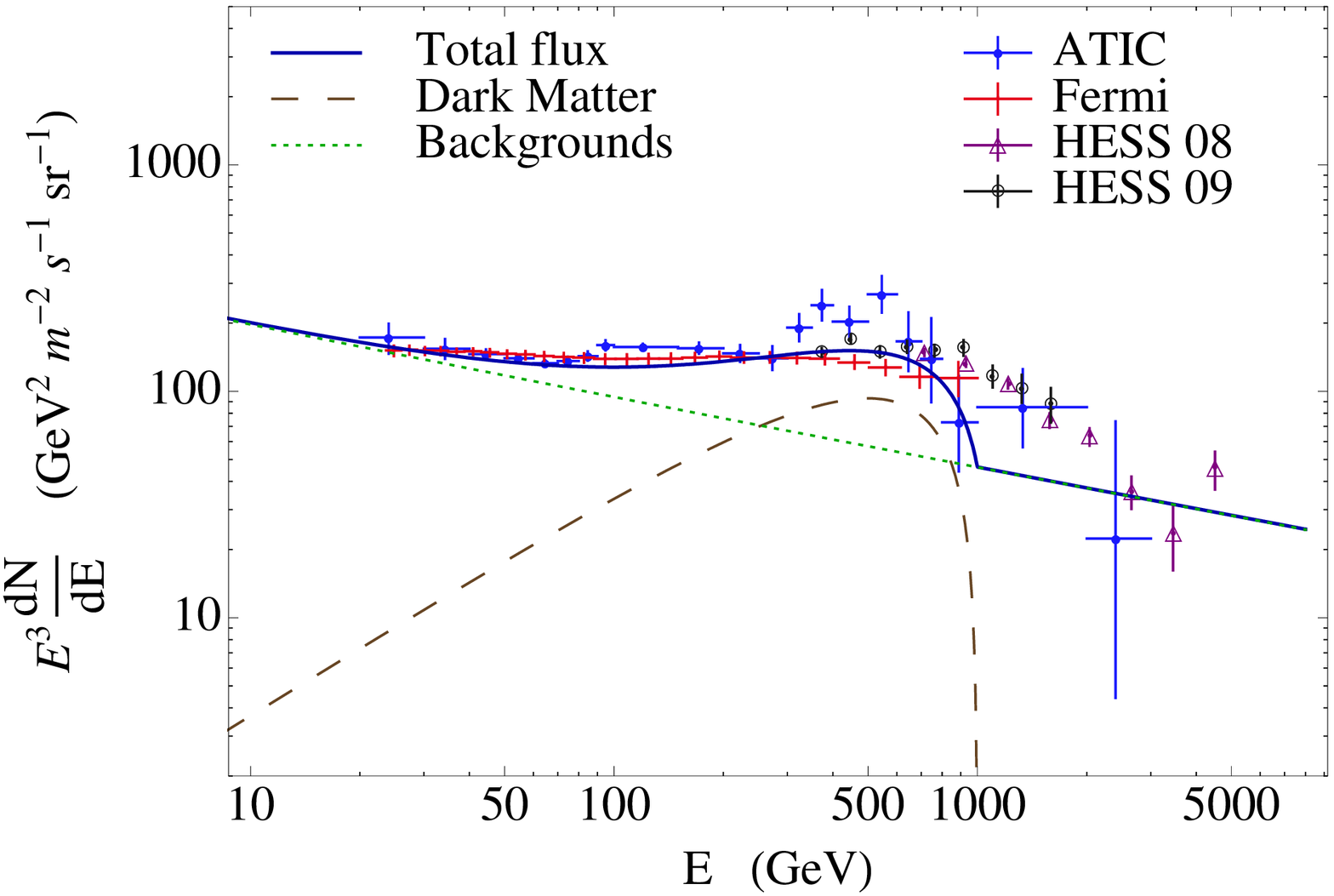,scale=0.4}
\hspace{3mm}
\epsfig{figure = 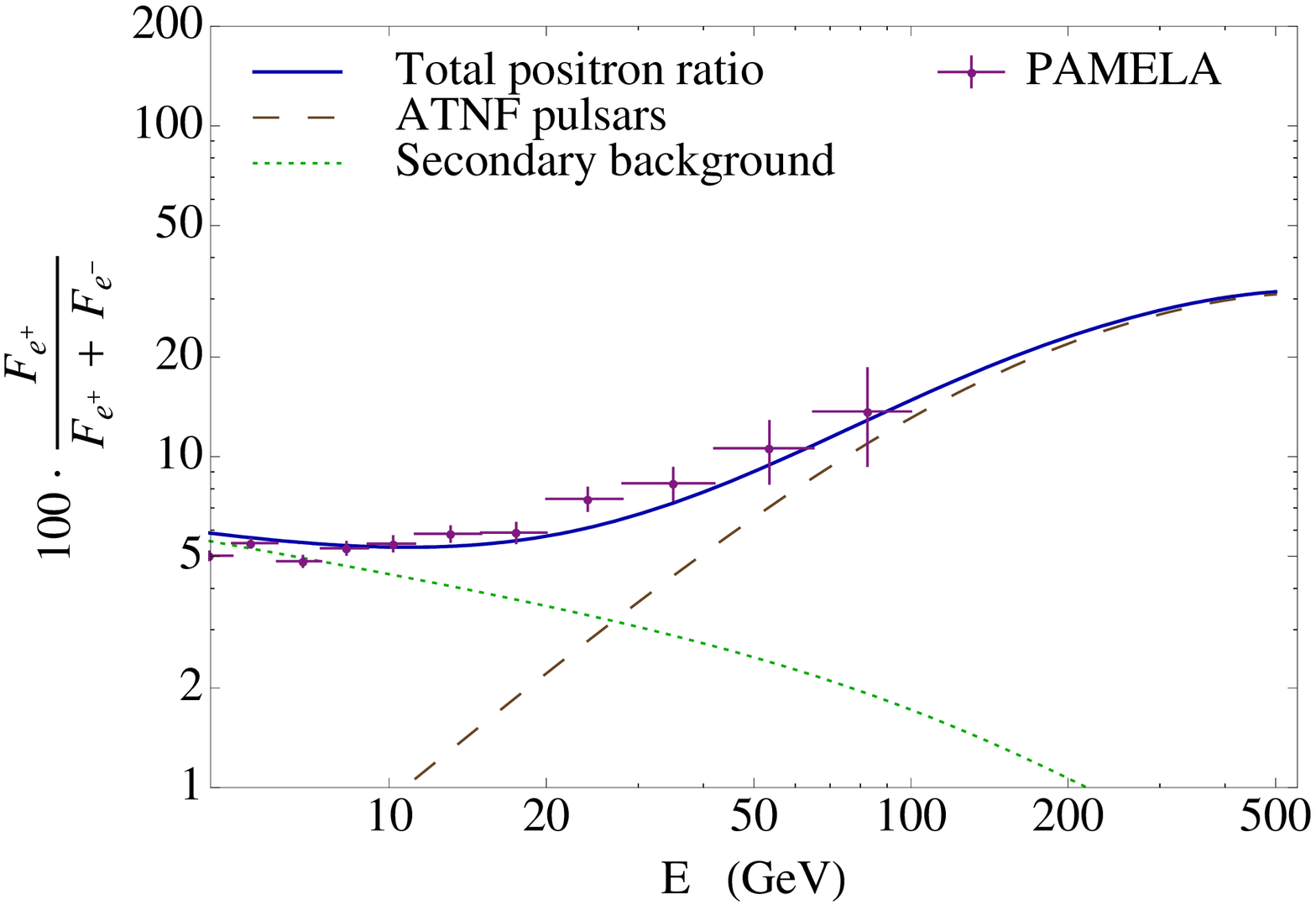,scale=0.4}
\vspace{-2mm}
\end{center}
\noindent
\caption{\small Flux from DM model in \cite{ArkaniHamed:2008qn} with
the annihilation chain $\chi + \chi \lra \phi + \phi \lra 2 e^+ + 2
e^-$, $M_{\rm DM} = 1$ TeV and a boost factor $BF \approx 500$.  The
primary background is $\sim E^{-3.3}$ and the secondary background is
$\sim E^{-3.6}$.  }
\label{DMplots}
\end{figure}

In this section we briefly review the $e^+e^-$ production from
annihilating (decaying) DM and derive that, for a large class of DM
models, the expected flux has the form of a power law with a universal
index $n = 2$ at energies $E \ll M_{\rm DM}$.  If we neglect gradients
in the DM density near the Earth, then we approximate any DM
contribution as originating from a constant, homogeneous source, which
from Eq.~(\ref{diff-loss}) gives 
\be
\rho(E) = \frac{1}{b(E)} \int_E^\infty Q(E') dE'.
\ee
This equation has an interesting
property that for any $Q(E) \sim E^k$ with $k > -1$, the integral is
saturated at the upper limit, which in this case is the mass of the DM
particle $M_{\rm DM}$.  For energies $E \ll M_{\rm DM}$, we can
neglect the dependence on $E$ resulting from the lower limit of
integration so the index of the electron flux is determined by the
index of the energy loss function $b(E) \sim E^2$.

The source function of $e^{+}e^{-}$ coming from annihilating dark
matter is \cite{Bertone:2004pz}
\be
\lb{DM_flux_annih}
Q (E) = \frac{1}{2} n_\chi^2 \bra \sigma v \ket \frac{dN}{dE},
\ee 
where
$n_\chi$ is the dark matter number density, $\langle\sigma v \rangle$
is the thermally averaged annihilation cross-section, and ${dN}/{dE}$
is the number density of electrons and positrons produced per
annihilation event.  Here we assume that the DM particle is its own
antiparticle otherwise there is an extra factor of $1/2$ in
Eq.~(\ref{DM_flux_annih}).  For this source function, the flux of
electrons and positrons from annihilating DM is 
\be
\lb{DM_annih}
F (E) = \frac{c}{8\pi}\frac{1}{b (E)} n_\chi^2 \bra \sigma v \ket  \int_E^M \frac{dN}{dE}(E') \, dE'.
\ee 
If the integral in this
equation is saturated at $E_* \lesssim M_{\rm DM}$, then for $E \ll
E_*$ the integral is insensitive to the changes of the lower
integration limit and can be approximated by a constant 
\be
\lb{DMmdlNorm} I_{e^\pm} = \int_0^M \frac{dN}{dE}(E') \, dE', 
\ee 
where $I_{e^\pm}$ is the average number of electrons and positrons produced
in an annihilation event.  In this case, the only energy dependence in
$F(E)$ is from $b(E)$, so $F(E) \sim E^{-2}$.  The discussion of the
universality of index $n = 2$ with respect to the choice of DM models
and DM halo profiles is further discussed in \cite{Kuhlen:2009is} (see
also \cite{Pohl:2008gm}\cite{Hooper:2008kv} for an earlier discussion
of the effects of DM substructure).

An important difference between the DM and pulsar models is that the
dark matter flux in Eq. (\ref{DM_annih}) has significantly fewer free
parameters than the corresponding flux from pulsars.  In fact, if we
assume that the energy losses in the ISM are well understood and the
energy density of dark matter is fixed from the cosmological
considerations, there are only two free parameters, $M_{\rm DM}$ and
$\bra \sm v \ket$, with the specific DM model providing $I_{e^\pm}$.
For a given DM model, $M_{\rm DM}$ is then fixed by the cutoff energy
in the observed spectrum and the cross section $\bra \sm v \ket$ is
fixed by the normalization of the flux.  The index of the flux is not
parametrically independent, $n \approx 2$.  This index is insensitive
to the choice of DM model or the DM profile in the host halo but may
change significantly in the presence of a large DM subhalo
\cite{Kuhlen:2009is}.  As an example, we use the DM model in
\cite{ArkaniHamed:2008qn} with the annihilation chain $\chi + \chi
\lra \phi + \phi \lra 2 e^+ + 2 e^-$.  DM with the current estimated
energy density of $\rho_\chi = 0.3\:\text{GeV} \text{cm}^{-3}$
requires $\langle\sigma v\rangle_0 = 3.0\times
10^{-26}\text{cm}^{3}\text{s}^{-1}$ at freeze out.  One can assume
that the current cross section is larger by a boost factor (BF).  To
fit the ATIC and PAMELA data, we set $M_{\rm DM} = 1$ TeV, which
requires a $BF \sim 500$ to reproduce the observed normalization
 (Fig.~\ref{DMplots}).

For a decaying DM model, Eq.~(\ref{DM_annih}) would be replaced by \be
\lb{DM_flux_decay} F (E) = \frac{c}{4\pi}\frac{1}{b (E)}
\frac{n_\chi}{\tau_{\rm d}} \int_E^M \frac{dN}{dE'} \, dE', \ee where
$\tau_{\rm d}$ is the life-time of the DM particle and $ \int
\frac{dN}{dE} \, dE$ is the number of electrons and positrons produced
per decay.  If we take the same number density and the mass of DM
particles as above, then 
\be 
\frac{I}{\tau_{\rm d}} \sim 5\times 10^{-27}\: \text{s}^{-1}.  
\ee 
These estimates agree with the analysis 
of \cite{Cholis:2008wq} \cite{Hisano:2008ah} \cite{Liu:2008ci}.


\section{Conclusions}
\lb{sec:conclusion}

In this work, we analyzed the flux of electrons and positrons from a
single pulsar, from a continuous distribution of pulsars, from pulsars
in the ATNF catalog and from dark matter.  Depending on the model
parameters and pulsar properties, they all can adequately fit either
the {\it Fermi} and {\it PAMELA} data or the ATIC and {\it PAMELA}
data.  One of the most important question is whether it is possible to
distinguish among these possibilities.

\begin{figure}[pt] 
\begin{center}

\epsfig{figure = 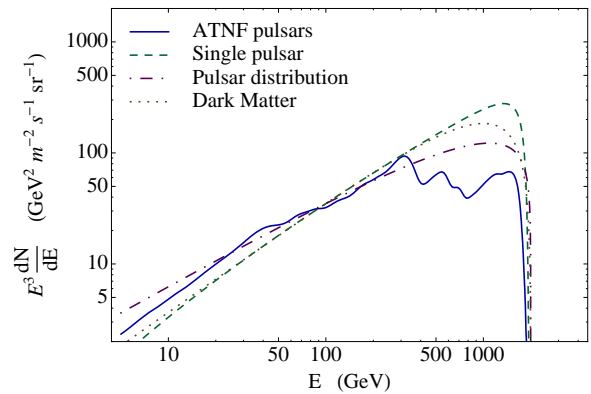,scale=0.4}
\end{center}
\noindent
\caption{\small The fluxes from annihilating dark matter, from a
single pulsar, and from a continuous distribution of pulsars can be
made similar, depending on the parameters of the models.  The flux
from a collection of pulsars may have significant deviations from a
continuous curve.  This property can be used to distinguish the
pulsars from the sources producing a featureless spectrum.  The flux
from pulsars in the ATNF catalog is the same as in Fig. \ref{PsrsFig}.
The single pulsar has the age t = 100 kyr, distance 0.3 kpc, $\eta W_0 =
9.2\times 10^{48}\:$erg, and $n = 1.6$.  The continuous pulsar distribution
has $N_{\rm b} = 1.8\:\text{kyr}^{-1}$, $\eta W_0 = 6.5\times
10^{48}\:$erg, $n = 1.5$, and $M_{\rm stat} = 2$ Tev.  The dark matter model
is the same as in Sec. \ref{sec:DM} but with $M_{\rm DM} = 2$ TeV and
$BF = 2000$.  The ISM properties are the same as in
Sec. \ref{sec:prop}.  }
\label{1similarity}
\end{figure}

In Fig. \ref{1similarity} we compare the expected $e^+e^-$ flux from a
single pulsar (Sec. \ref{sec:single}), pulsars in the ATNF catalog
(Sec. \ref{sec:distr}), a continuous distribution of pulsars
(Sec. \ref{sec:distr}), and DM (Sec. \ref{sec:DM}).  We have chosen
the parameters of the models such that the fluxes have the same value
at 100 GeV, similar indices at low energies, and a cutoff at 2 TeV.
At energies below $\sim 300$ GeV, the fluxes are very similar.  We
also do not expect to see any differences between these models in the
positron ratio below 300 GeV, the upper limit for charge
identification in {\it PAMELA}.

Above 300 GeV, there are substantial differences among the $e^+e^-$
spectrum predicted for these models.  However, it should be noted that
the sharpness of the cutoff for a single pulsar and for DM is strongly
model dependent.  If the injection cutoff for a pulsar is $\sim 1$
TeV, then the cutoff in the observed flux from a single pulsar can be
much smoother than if the injection cutoff was higher than the cooling
break, in which case its spectrum is indistinguishable from that
predicted for a continuous distribution of pulsars.  For the DM flux
we show a model with only one intermediate particle in the
annihilation-decay process.  If there are more steps in the
annihilation-decay process, then the flux has a broader cutoff and, again, may
be impossible to distinguish between either a single pulsar or
continuous pulsar distribution origin. Thus, given the significant
uncertainties in the pulsar and DM models, it is unlikely that better
observations alone can distinguish between a single pulsar and dark
matter origin of anomalous $e^+e^-$ flux \cite{Hall:2008qu} (a similar
conclusion was obtained in \cite{Ioka:2008cv}\cite{Zhang:2008tb}).

The flux from a discrete collection of pulsars does have a few
distinctive features at high energies.  The height of these features
is model dependent and may be within the error bars of current
observations.  The presence of these features requires the existence
of a few young, nearby, energetic pulsars with an injection cutoff
$\gg 1$ TeV.  Consequently the absence of such features in the
observed $e^+e^-$ spectrum could mean that all young pulsars whose
electrons have reached the Earth have had cutoffs $\lesssim 1$ TeV --
a strong constraint on the properties of PWNe since, as we discuss in
Appendix \ref{Pulsars-app}, the PWN around the Vela pulsar has a cutoff in
the electron and positron spectrum at an energy $\gg 1$ TeV.  
An additional smearing of the bumps can be due to spatial variations of the 
energy losses and the diffusion coefficient.
Our general conclusion is that the current electron and positron data are not
sufficient to distinguish between the pulsars and the dark matter and
that independent measurements, e.g., the spectrum and morphology of the
diffuse galactic gamma-ray background \cite{Zhang:2008tb,
Ando:2009fp}, may be necessary in order to decisively distinguish a
pulsar and DM origin of $e^+e^-$ excess.

\bigskip
\bigskip
\bigskip

\noindent
{\large \bf Acknowledgments.}
\noindent
The authors are thankful
Gregory Gabadadze, Andrei Gruzinov, Ignacy Sawicki, 
Jonathan Roberts, and Alex Vikman
for valuable discussions.
We are especially indebted to Neal Weiner for initiating the project
and for numerous discussions and support during all stages of the
work.  This work is supported in part by the Russian Foundation of
Basic Research under Grant No. RFBR 09-02-00253 (DM), by the NSF Grants No.
PHY-0245068 (DM) and No. PHY-0758032 (DM), by DOE OJI Grant No. DE-FG02-06E
R41417 (IC), by the NSF Astronomy and Astrophysics Postdoctoral
Fellowship under Grant No. AST-0702957 (JG).

\newpage

\appendix


\section{Review of pulsars}
\lb{Pulsars-app}

In this appendix we review the emission of electrons from pulsars.  We
assume that this emission is powered by the pulsar's loss of
rotational energy.  Pulsars are believed to be rotating neutron stars
with a strong surface magnetic field \cite{Shapiro1983}, and magnetic
dipole radiation is believed to provide a good description for its
observed loss of rotational energy.  A pulsar loses its rotational
energy on a characteristic decay time $\tau$ defined as 
\be 
\tau = \frac{\mcE_0}{\dot \mcE_0}, 
\ee 
where ${\mcE_0}$ and ${\dot \mcE_0}$
are the initial rotational energy and the initial spin-down
luminosity, respectively, which in the magnetic dipole radiation model are equal to
\beaa 
\mcE_0 &=& \frac{1}{2} I \Om_0^2, \\ \dot \mcE_0 &=& \frac{B^2 R^6 \sin^2\al}{ 6c^3 } \Om_0^4, 
\eeaa 
where $\Om_0$ is the initial
angular velocity, $R$ is the radius of the pulsar, $B$ is the strength
of the surface dipole magnetic field, and $\al$ is the angle between
the rotation axis and the magnetic field axis.

If the energy loss is due to magnetic dipole radiation, then 
\be \lb{m-dipol} 
I \Om \dot\Om = - \frac{B^2 R^6 \sin^2\al}{ 6c^3 } \Om^4.
\ee 
Integrating the energy loss equation we get 
\bea \Om(t) &=& \Om_0 \left(1 + \frac{t}{\tau}\right)^{-\frac{1}{2}}, \\ 
\lb{EdotEvolution}
\dot{\mcE}(t)&=& \mcE_0 \frac{1}{\tau}\left(1 + \frac{t}{\tau}\right)^{-2}.  
\eea 
As a result, the pulsar angular
velocity satisfies 
\be \lb{psr-tau} 
\frac{\Om}{2\dot{\Om}} = -(t + \tau).  
\ee 
In a more general approach, the time evolution of the
angular velocity is described as 
\be 
\dot\Om \sim - \Om^k, 
\ee 
where $k$ is the breaking index, which can be found by measuring the current
$\Om$, $\dot\Om$, and $\ddot\Om$ 
\be 
k = - \frac{\Om\ddot\Om}{\dot\Om^2}.  
\ee 
In this case, 
\be \Om(t) = \Om_0 \left(1 + \frac{t}{\tau}\right)^{-\frac{1}{k - 1}}.  
\ee 
The magnetic dipole radiation corresponds to $k = 3$.

As an example, let us calculate the initial rotational energy of the
Crab pulsar using the magnetic dipole approximation and a general
braking index.  The Crab pulsar is believed to have been produced
during SN 1054 supernova explosion.  Consequently, the age of the
pulsar is known exactly, $t = 955$ yr.  In the magnetic dipole
approximation, we can use Eq. (\ref{psr-tau}) and the current values
of $\Om$ and $\dot\Om$ \cite{Manchester:2004bp, 1993MNRAS.265.1003L} to calculate
that its pulsar time scale $\tau \approx 0.3$ kyr.  Assuming a mass $
1.4\, M_\odot$, radius $R = 12$ km, and moment of inertia $I =
1.4\times 10^{45}\: \text{g} \: \text{cm}^2$ \cite{Shapiro1983}, we
derive an initial rotational energy $W_0 \approx 3\times
10^{50}$\:erg.  Taking into account the measured value of $\ddot\Om$
\cite{Manchester:2004bp}, the braking index of the Crab pulsar is $k =
2.5$, which gives $\tau \approx 0.7$ kyr and $W_0 \approx 5.3\times
10^{49}\:$erg.

\begin{figure}[tp] 
\begin{center}
\epsfig{figure = 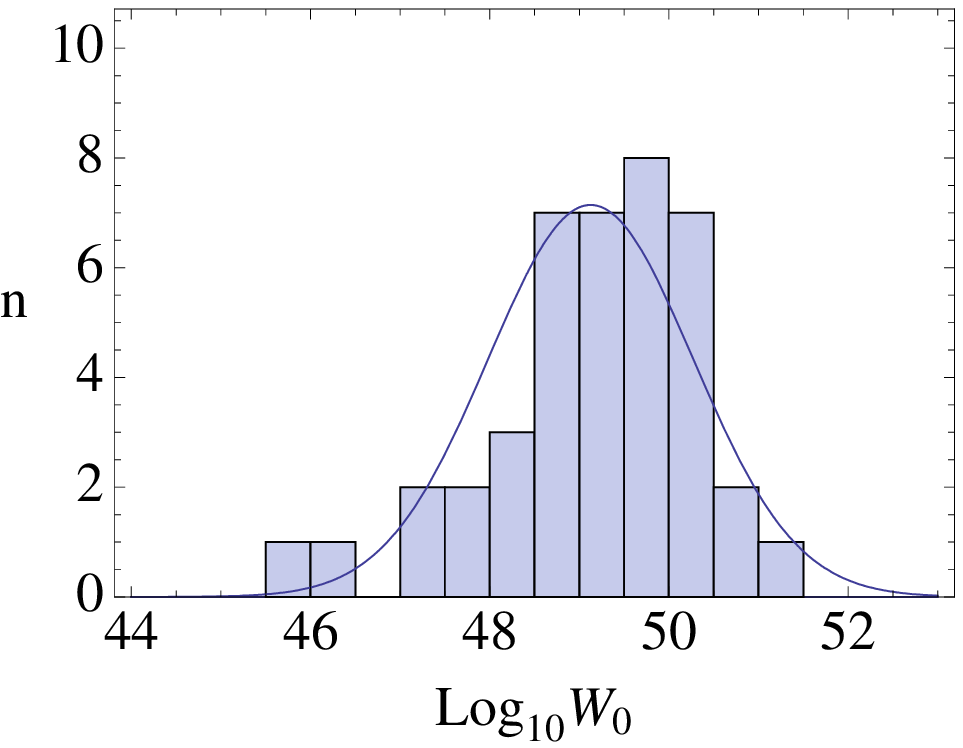,scale=.42}
\epsfig{figure = 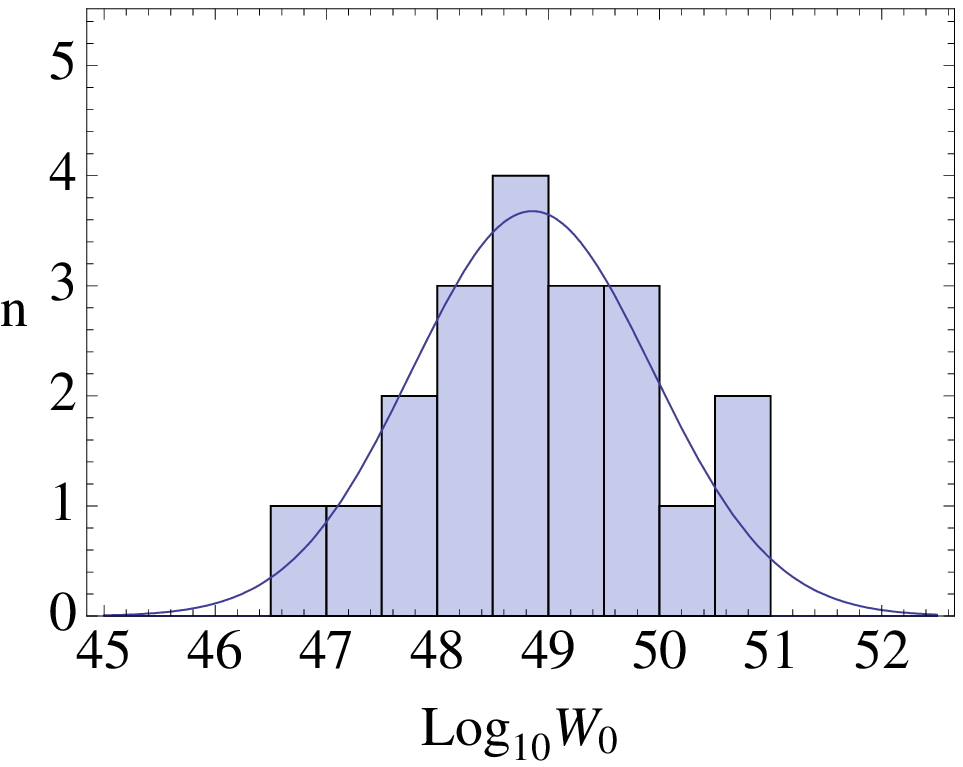,scale=.42}
\vspace{-2mm}
\end{center}
\noindent
\caption{\small The left panel has 41 pulsars within 4 kpc from the Earth
and younger than 300 kyr.  Assuming the braking index 3 and the pulsar
time scale $\tau = 1$ kyr we find the average power $p = {\rm
Log}_{10}(W_0/\text{erg}) = 49.1 \pm 1.1$ with $\chi^2/dof = 1.2$.
For the right panel we select the pulsars from the ATNF catalog that
have braking index $2< k < 10$ (there are 20 such pulsars).  Assuming
the pulsar time scale $\tau = 1$ kyr, these pulsars have $p = 48.9 \pm
1.1$ with $\chi^2/dof = 0.3$.  For the pulsar time scale $\tau = 10$
kyr, the left (right) selection of pulsars would have $p = 48.2 \pm
1.1 (48.1 \pm 1.2)$.  Thus, for $\tau = $ 1 kyr (10 kyr) the average
initial rotational energy is $\bar W_0 \approx 10^{50}\:$erg
($10^{49}\:$erg).  }
\label{W0calculation}
\end{figure}

If the age of the pulsar is not known independently, it is impossible
to determine $\tau$ and $W_0$ using its observed properties (from
$\Omega$, $\dot\Om$, and $\ddot\Om$ one can only calculate $t +
\tau$).  In the following, we estimate $W_0$ by assuming $\tau = 1$
kyr for all pulsars.  If so, the initial energy $W_0 \equiv \mcE_0$
can be found from Eq. (\ref{EdotEvolution}) by using the current
spin-down luminosity $\dot\mcE$ 
\be \lb{mdipW0} 
\mcE_0 = \dot\mcE \tau (1 + t/\tau)^2.
\ee 
For a general braking index $k$, this formula takes
the form 
\be \lb{brIndW0} 
\mcE_0 = \dot\mcE \tau (1 + t/\tau)^\frac{k + 1}{k - 1} 
\ee 
Using this method to estimate $W_0$ for all pulsars in
the ATNF catalog results in the distribution shown in
Fig. \ref{W0calculation}, using both the magnetic dipole approximation
(Eq. (\ref{mdipW0})) and a general braking index method
(Eq. (\ref{brIndW0})).  In the magnetic dipole case we took all
pulsars within 4 kpc from the Earth and younger than 300 kyr.  The
reason is that older and more distant pulsars are less luminous and
may not be observed for small spin-down luminosities, i.e., this
introduces a bias towards more energetic pulsars and shifts the
distribution toward larger average $W_0$.  For the general braking
index, we used all pulsars with $2 < k < 10$.  In both cases, $W_0$
can be described by a log-normal distribution with the average $p =
{\rm Log}_{10}(W_0/\text{erg}) \approx 49$ and the standard deviation
$\sm_p \approx 1$.  The average initial rotational energy in this case is 
\be 
\bar W_0 = \frac{1}{\sqrt{2\pi} \sm_p}\int 10^p 
e^{-\frac{(p - \bar p)^2}{2\sm_p^2}}dp 
\approx 10^{50}\:\text{erg}. 
\ee 
This result is strongly dependent on the chosen value of $\tau$.  
For example, if $\tau = 10$ kyr, then an analogous calculation gives 
$\bar p \approx 48$, $\sm_p \approx 1$ and $\bar W_0 \approx 10^{49}\:$erg.  
The estimations above agree with the analysis of \cite{FaucherGiguere:2005ny}. 
It is worth noting that $\tau$ likely varies between pulsars.

To estimate $\eta$, it is important to understand how the pulsar's
magnetic dipole radiation is transferred to the kinetic energy of
particles.  Since $\dot\mcE$ decays as $t^{-2}$, most of the
rotational energy is lost at early times.  A young pulsar is
surrounded by several layers \cite{Rees:1974nr}\cite{Kennel:1984vf}.
Nearest to the neutron star is the magnetosphere, which ends at the
light cylinder $R_{\rm LC} \equiv c / \Om$.  The rotating magnetic
field creates a strong electric field capable of both producing pairs
of particles and accelerating them to relativistic energies.  These
particles stream away from the light cylinder as a coherent ``wind''
that ends with a termination shock separating the wind zone from
the PWN which consists of magnetic fields and
particles moving in random directions.  The PWN in turn is surrounded
by an SNR.  A significant PWN exists only at the
early times ($t \ll 100$ kyr \cite{Gaensler:2006ua}).

The spectrum of electrons and positrons in the magnetosphere can be
estimated using the spectrum of pulsed $\g$-ray emission from a
pulsar.  For the Crab pulsar, model fits to the observed photon
spectrum suggest that the spectrum of $e^{+}e^{-}$ pairs in its
magnetosphere is well described by a broken power low with an index of
$2.0$ below $E_{\rm br} \sim 2$ GeV and an index of $2.8$ between $E_{\rm
br}$ and an upper cutoff around $100$ GeV \cite{Harding:2008kk} (see
also \cite{Aliu:2008hc}).  Particles with this spectrum {\it cannot}
reproduce the $e^+e^-$ spectrum observed on Earth, since a break in
the injection spectrum at $2$ GeV is too low to explain the ATIC and
{\it PAMELA} results, and an index of $2.8$ above 2 GeV also does not
fit the data.  Additionally, the pulsed emission from a pulsar only
reflects the energy spectrum of the emitting particles in the emission
region, which is not necessarily representative of the spectrum of
particles that escape the pulsar magnetosphere along the open field
lines and are eventually deposited in the ISM.

These particles are further accelerated before they enter the PWN,
most likely at the termination shock between the magnetosphere and the
PWN (for a review see, e.g., \cite{Arons1996}).  Once deposited in the
PWN, they are trapped by the PWN's magnetic field until it is
disrupted.  Observationally, the spectrum of the electrons inside the
PWN is found by analyzing their broadband spectrum, which at low
photon energies ($<1$ GeV) is dominated by synchrotron emission and at
higher energies ($>100$ GeV) dominated by inverse Compton scattering
of electrons off background photons \cite{1996MNRAS.278..525A}.  From the
radio spectrum of these objects, it is possible to constrain the
spectral shape of the low energy (GeV) electrons which dominate by
number the electron population of a PWN.  An average index of the
electron and positron spectrum can be found using data from the
publicly available {\it Catalogue of galactic SNRs} \cite{Green:2009qf},
where F-type (or ``filled-center") SNRs are PWNe, S-type are the
supernova shells, and C-type SNRs are a combination of the two.
There are 7 F-type SNRs with an average electron index $n_{\rm F}
\equiv 2\al + 1 = 1.3 \pm 0.3$.  For 21 C-type SNRs the average index
is $n_{\rm C} = 1.8 \pm 0.4$, while 168 S-type SNRs have $n_{\rm S} =
2.0 \pm 0.3$.  It is clear that the spectrum of electrons in
supernova shells is much softer (decreases faster with the energy)
than the spectrum in PWNe. 
In order to explain the {\it PAMELA} positron ratio
we need either F-type or, possibly, C-type SNRs
because the S-type SNRs are produced by the initial supernova explosion
and do not contain a significant number of positrons.


The broadband spectrum of most PWNe shows a break between the radio
and X-ray regimes, believed to correspond to a break in the electron
and positron spectrum, most likely the result of synchrotron cooling.
Converting the frequency of this break to an electron/positron energy
requires knowing the strength of the PWN's magnetic field.  An
independent estimate of the magnetic field is available for those PWNe
with detected inverse Compton emission, since this depends solely on
the energy spectrum of electrons and positrons in the PWN and known
properties of the various background photon fields (e.g., Cosmic
Microwave Background and starlight).  The best studied example is the
Crab Nebula (e.g., \cite{1996MNRAS.278..525A}), whose broadband photon
spectrum suggests an electron spectrum well described by a broken
power law with an index $n = 1.5$ below $E_0 \sim 200$ GeV and $n =
2.4$ between $E_0$ and an upper cutoff $E_{\rm cut} \sim 10^3$ TeV:
the magnetic field in this PWN has a strength of $B \approx 2\times
10^{-4}$ G, resulting in a ratio of magnetic energy flux to particle
energy flux of $\sm < 0.01$ \cite{1996MNRAS.278..525A}.  For PWNe whose
broadband spectrum is not as well determined, the break energy in the
electron spectrum is typically derived using the minimum energy
assumption ($\sm=0.75$ \cite{Chevalier:2004rp}).  Using this method
and the observational data provided in \cite{Chevalier:2004rp}, we
estimate a break energy of $\sim3-300$ GeV for the PWNe listed in this
paper.  It is important to emphasize that this procedure almost
certainly overestimates the magnetic field strength inside a PWN since,
for most PWNe, $\sm$ is believed to be $\sm\ll0.75$.  In this
procedure, the inferred break energy $E_b$ is $E_b \propto B_{\rm
pwn}^{-1/2}$, where $B_{\rm pwn}$ is the strength of PWN's magnetic field.
As a result, the true break energy of electrons and positrons inside
the PWNe analyzed above is likely to be at least an order of magnitude
higher than the derived value.

The break energy in the electron/positron spectrum of a PWN is
expected to vary considerably during the lifetime of a PWN due largely
to changes in the strength of the PWN's magnetic field
(e.g., \cite{Reynolds:1984}, \cite{Gelfand:2009aa}).  Therefore, the
break energy in the spectrum of electrons and positrons injected by
the PWN into the surrounding ISM depends strongly on the evolutionary
phase of the PWN when this occurs.  During the initial free-expansion
phase of the PWN's evolution (the Crab Nebula is the prototypical
example of such a PWN; \cite{Gaensler:2006ua}), the break energy is
expected to increase as $\sim t^{1.6}$ \cite{Reynolds:1984,
Gelfand:2009aa}.  This phase of the PWN's evolution ends when it
collides with the SNR's reverse shock, typically on the order of
$\sim10^4$ yr after the supernova explosion.  If this holds for the
Crab Nebula, its current age of $\sim1000$ yr and break energy of
$\sim 200$ GeV suggests that, at the time of this collision, the break
energy will have risen to $\sim 8$ TeV.  The evolution of the break
energy after this collision is more complicated and depends strongly
on the properties of the central neutron star, progenitor supernova,
and surrounding ISM (see \cite{Gelfand:2009aa} for a more detailed
discussion).  There is observational evidence that the break energy of
older PWNe ($>10^4$ yr old) is considerably higher than that of the
Crab Nebula and other young PWNe ($\sim10^3$ yr).  The most
convincing example comes from a recent analysis of the broadband
spectrum (radio to TeV $\gamma$ rays) of the Vela PWN (often referred
to as ``Vela X''), which suggests a break in the electron spectrum of
$\sim67$ TeV \cite{Aharonian:2006xx}.  For the purpose of the work
presented here, the exact value of the break is not important as far
as it is bigger than $\approx 1$ TeV.

Observations indicate that most PWNe are particle dominated:
i.e., almost $100\%$ of the spin-down luminosity is transformed into
the energy of the particles after the termination shock.  However, not
all of this energy is eventually deposited in the ISM.  We estimate
this fraction, $\eta$, by first assuming that the spectrum after the
termination shock is a power law $Q (E) \sim E^{-n}$ with an index $n
< 2$ and a cutoff $E_{\rm c} \sim 10^3$ TeV.  Then, the total energy
in electrons is \be W_{\rm ini} \sim \int E^{-n} EdE \sim E_{\rm c}^{2
- n}.  \ee If, when the PWN is disrupted, the energy spectrum of
electrons in the PWN is $E^{-n}$ below the break at $E_{\rm br}$ and
$E^{-n_{\rm b}}$ with $n_{\rm b} > 2$ above the break, the total
energy in electrons is saturated at $E_{\rm br}$ with \be W_{\rm fin}
\sim E_{\rm br}^{2 - n}.  \ee The efficiency is therefore \be \eta =
\frac{W_{\rm ini}}{W_{\rm fin}} \sim \left( \frac{E_{\rm br}}{E_{\rm
c}} \right)^{2 - n}.  \ee For $n = 1.5$, $E_{\rm br} = 10$ TeV and
$E_{\rm c} = 10^3$ TeV, this gives the suggested $\eta = 0.1$.  This
derivation should be viewed as an order of magnitude estimation.  A
more realistic calculation is extremely complicated and involves the
knowledge of the PWN evolution and the actual spectra of particles
inside a PWN.  Recent work in this field does support an efficiency of
$\eta\sim 0.1$ (e.g., \cite{Gelfand:2009aa}).

It should be stressed that, apart from theoretical uncertainties, the
parameters of the injection spectrum can vary significantly between
pulsars.  The initial rotational energy can differ by several orders
of magnitude, while the index of the electron spectrum $n$ can vary
from $1$ to $2$.  In some cases, it is observed to vary inside the PWN
of a single pulsar. 
The upper cutoff $M$, as we have seen in the examples of Crab 
and Vela pulsars, can vary at least between $\sim\! 100$ GeV and $\sim\! 10$ TeV.  



\section{Spatial variation in energy losses}
\lb{Smearing-app}

As we have discussed, an important signature of the flux from the
pulsars is the presence of a number of bumps at the cooling break
energies 
\be
E_i = \frac{1}{b t_i},
\ee
 where $t_i$'s are the ages of
the pulsars.  The existence of these bumps is based on the assumption
that the energy losses depend only on the travel time and not their
path.  In reality, the energy loss coefficient depends on the
position, since the densities of the star light and IR photons vary in
space.  In this case, there is no simple solution for
Eq.~(\ref{diff-loss}), though one can still find the average energy
loss and its standard deviation by averaging the energy losses over
random paths.

As a useful simplification we will consider separately diffusion in
space and energy losses.  Our motivation is that a particle detected
with energy $E$ has an energy close to $E$ during most of the
propagation time (i.e. the cooling time from $E_0$ to $E$ is saturated
by the final energy $E$).  Consequently, the diffusion coefficient for
all particles detected with energy $E$ can be approximated by $D(E)$.
If so, the probability to propagate from a source at $(x_0,\: t_0)$ to
an observer at $(x_1,\: t_1)$ is given by the Green function
\be\lb{simpleDiffGreen}
G(x_1,\: t_1;\: x_0,\: t_0) = \frac{1}{(4\pi D(E) \Dl t)^{3/2}}
e^{-\frac{ \Dl \vx^2}{4 D(E)\Dl t}}.
\ee
In order to find the energy loss averaged over paths, it is useful to rewrite
this Green function in terms of the path integral 
\be
G(\vx_1,\: t_1;\: \vx_0,\: t_0) = \int D x(t) \; e^{- S[x(t)]},
\ee
where the
action is 
\be
S[x(t)] = \int \frac{1}{4D(E)}{\dot x}^2 dt
\ee
with the boundary conditions $x(t_0) = x_0$ and $x(t_1) = x_1$.

In general, the average of a functional $\OO[x(t)]$ over paths is
\be\lb{avFncl}
\bra \OO \ket = \frac{ \int D x(t) \;\OO[x(t)]\; e^{- S[x(t)]} }{  \int D x(t)\; e^{- S[x(t)]} }.
\ee
Integrating the energy loss
\be
\frac {dE}{dt} = - b(x) E^2
\ee
along a path $x(t)$, we find
\be
\lb{ener-diff}
\frac {1}{E_1} - \frac {1}{E_0}  = \int_{t_0}^{t_1} b(x) dt.
\ee
The functional that we will study is 
\be \lb{lossFunctional}
\OO[x(t)] = \int_{t_0}^{t_1} b(x) dt.
\ee
The expression in the numerator of (\ref{avFncl}) is 
\bea
{\rm Num} & = & \int D x(t) \; \int_{t_0}^{t_1} dt'\: b(x(t')) \; e^{- S[x(t)]}\nonumber \\
& = & \int_{t_0}^{t_1} dt' \int D x(t) \;b(x(t')) \; e^{- S[x(t)]}.
\eea
If we define $x' = x(t')$, then all the paths
can be represented as a path from $x_0$ to $x'$, the integral over all
$x'$ and the path from $x'$ to $x_1$:
\bea
\nonumber
{\rm Num}
& = & \int_{t_0}^{t_1} dt' \int_{x_0}^{x'} D x(t) \int dx' \int_{x'}^{x_1} D x(t)\\
& & \cdot b(x(t')) \; e^{- S[x(t)]} \\
\lb{energyLoss}
& = & \int_{t_0}^{t_1} dt' \int dx'\; G(x_1,\: t_1;\: x',\: t')\nonumber \\
& & \cdot b(x')\; G(x',\: t';\: x_0,\: t_0) .
\eea
The resulting
expression resembles the first order perturbation theory: there is a
propagation from $x_0$ to $x'$, an insertion of an operator at $x'$ and a
propagation from $x'$ to $x_1$.

The average energy loss can be estimated with (\ref{energyLoss}) and
the Green function in (\ref{simpleDiffGreen}).  Taking $E_0 \ra
\infty$ in (\ref{ener-diff}), we find that the average cooling break
energy for a given pulsar is 
\bea
\lb{averBreak}
\bra \frac {1}{E_{\rm br}} \ket & \equiv & \bra \OO \ket  =  
  \text{\Large (}\; \int_{t_0}^{t_1} dt' \int dx'\; G(x_1,\: t_1;\: x',\: t')\; b(x')\nonumber\\
& & \cdot G(x',\: t';\: x_0,\: t_0)\;\text{\Large )}/
         G(x_1,\: t_1;\: x_0,\: t_0) .
\eea

The standard deviation is 
\be
\sm_\OO = \sqrt{\bra\OO^2\ket - \bra\OO\ket^2 }.
\ee
The average $\bra \OO^2 \ket$ for the functional
(\ref{lossFunctional}) can be computed analogously to
(\ref{averBreak}) 
\beaa
\bra \OO^2 \ket &=& 2\; G^{-1}(x_1;\: x_0)  \\
&& \int_{t_0}^{t_1} dt' \int dx'\; \int_{t'}^{t_1} dt'' \int dx''\; 
                G(x_1;\: x'')\\
&& \cdot b(x'')\;  G(x'';\: x')\; b(x') \;G(x';\: x_0),
\eeaa
where we assume that $x'$ is at $t'$ and $x''$ is at $t''$.  The
factor of 2 is the usual $n!$ for the time-ordered path integrals.

The relative standard deviation of the cooling break energy is 
\be
\frac {\Dl E}{E} = \frac {\sm_\OO}{\bra \OO \ket}.
\ee
Using the energy densities of starlight and IR photons from
\cite{Porter:2005qx} we find the relative smearing in the energy 
\be
\lb{eSmear}
\frac {\Dl E}{E} \approx 0.053 \cdot \left(\frac{E}{1\: TeV}\right)^{-1/3}.
\ee
At 1 TeV the smearing is about 5\%, at 100
GeV it is 11\%, and at 10 GeV it is 24\%.  The flux from the ATNF pulsars
with this smearing is shown in Fig.  \ref{PsrsFig}.  At low energies
the flux becomes very smooth but at high energies the bumps are still
visible.  We also notice that at high energies the relative width of
the bumps is larger than the ratio in (\ref{eSmear}).  Thus, even if
the experimental energy resolution is about 10\% -- 15\%, we should be
able to see the bumps.


\section{Constraining pulsars and ISM properties}
\lb{constr-app}

If we assume that the anomalies in {\it Fermi}, ATIC and {\it PAMELA}
data are due to pulsars, we can use these results to constrain the
properties of ISM and pulsars.  The problem is that the flux depends
on both the properties of the ISM and the injection spectrum from
pulsars.  As we discuss in Secs. \ref{sec:single} and
\ref{sec:distr}, the ISM can be described by three parameters $D_0$,
$\dl$, and $b_0$ and the injection from a distribution of pulsars can be
described by five parameters $W_0$, $\eta$, $n$, $M$, and $N_b$.
Obviously, the three parameters of the observed flux (the
normalization, the index, and the cutoff) cannot constrain all eight
parameters, but they can constrain some combinations of parameters.
These constraints may be very useful if combined with results from
other experiments, such as the observations of protons, heavy nuclei,
or diffuse gamma rays.

Another concern is the reliability of constraints coming from the
local $e^+e^-$ flux.  Ideally, we would like to constrain the
parameters in the models, but since the local distribution of pulsars
is fundamentally random in nature, there is a possibility that we can
only constrain the properties of particular pulsars without getting
any information about the general population.  The reason why we think
our approach is sensible is the following.  At high energies, the flux
from pulsars will depend significantly on the properties of individual
pulsars (and we can use this region to prove that the observed flux is
due to pulsars), but at low energies the flux is well approximated by
the continuous distribution flux and the properties of individual
pulsars are relatively unimportant.  Thus, we propose using the
observed spectrum at intermediate energies $100\: \text{GeV} < E <
500\: \text{GeV}$ as a testing ground to study the general (or
averaged) properties of $e^+e^-$ injection from pulsars.

In the following we will fit the continuous distribution flux derived
in Eq.~(\ref{ContDistrScaling}) to the {\it Fermi} and {\it PAMELA}
data simultaneously.  In these fits we substitute the usual pulsar
injection cutoff $M \sim 10$ TeV by the propagated (or statistical)
cutoff $M_{\rm stat}$ and treat it as a fit parameter.  The other fit
parameters are $b_{0} = (0.5-3)\times10^{-16}\:
\text{GeV}^{-1}\:\text{s}^{-1}$ (these values are based on the energy
densities of radiation and magnetic field within few kpc from the
Earth), $\dl = (0.3-0.6)$ \cite{Strong:2007nh}.  The injection index
$n \approx (1 - 2) $ and the conversion efficiency $\eta W_0 \sim
10^{48} - 10^{49}\:$erg are discussed in Appendix~\apref{Pulsars-app}.  In
the fits we use $D_0 = 100\: \text{pc}^2\text{kyr}^{-1}$, $N_{\rm b} =
1.8\: {\rm kyr}^{-1}$ and $R_{gal} = 20\: \text{kpc}$.  The best-fit
parameters are: $\delta=0.48$, $b_{0}$ =
$1.05\times10^{-16}\:\text{GeV}^{-1}\: \text{s}^{-1}$, $\eta
W_{0}$=$4.9\times10^{48}$ erg, $n=1.7$, and $M_{\rm stat}=1.0$ TeV.
The corresponding fluxes are shown in Fig. \ref{fig:ATIC_PAMELA_fits}.

\begin{figure*}[h]
\begin{center}
\includegraphics[scale = 0.6]{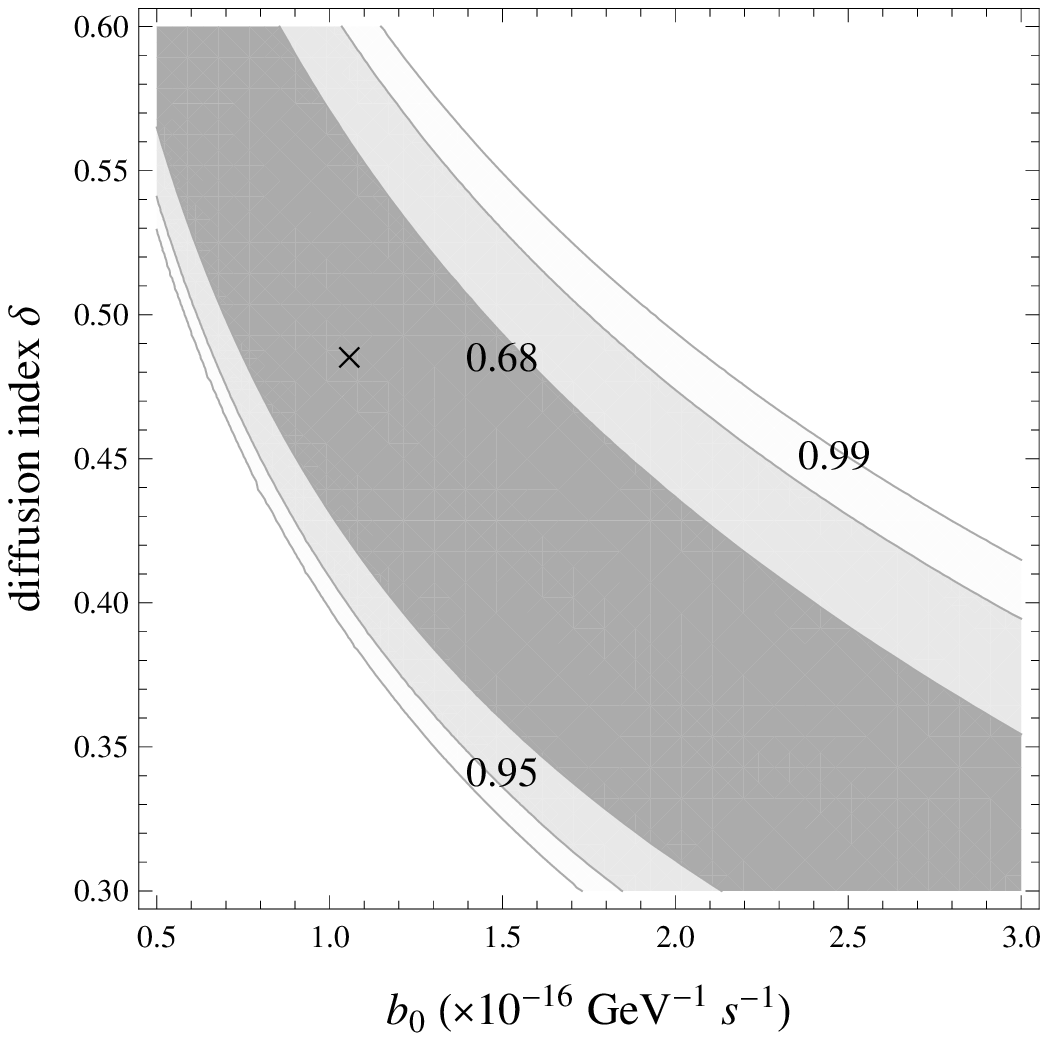}
\hskip 0.15in
\includegraphics[scale = 0.6]{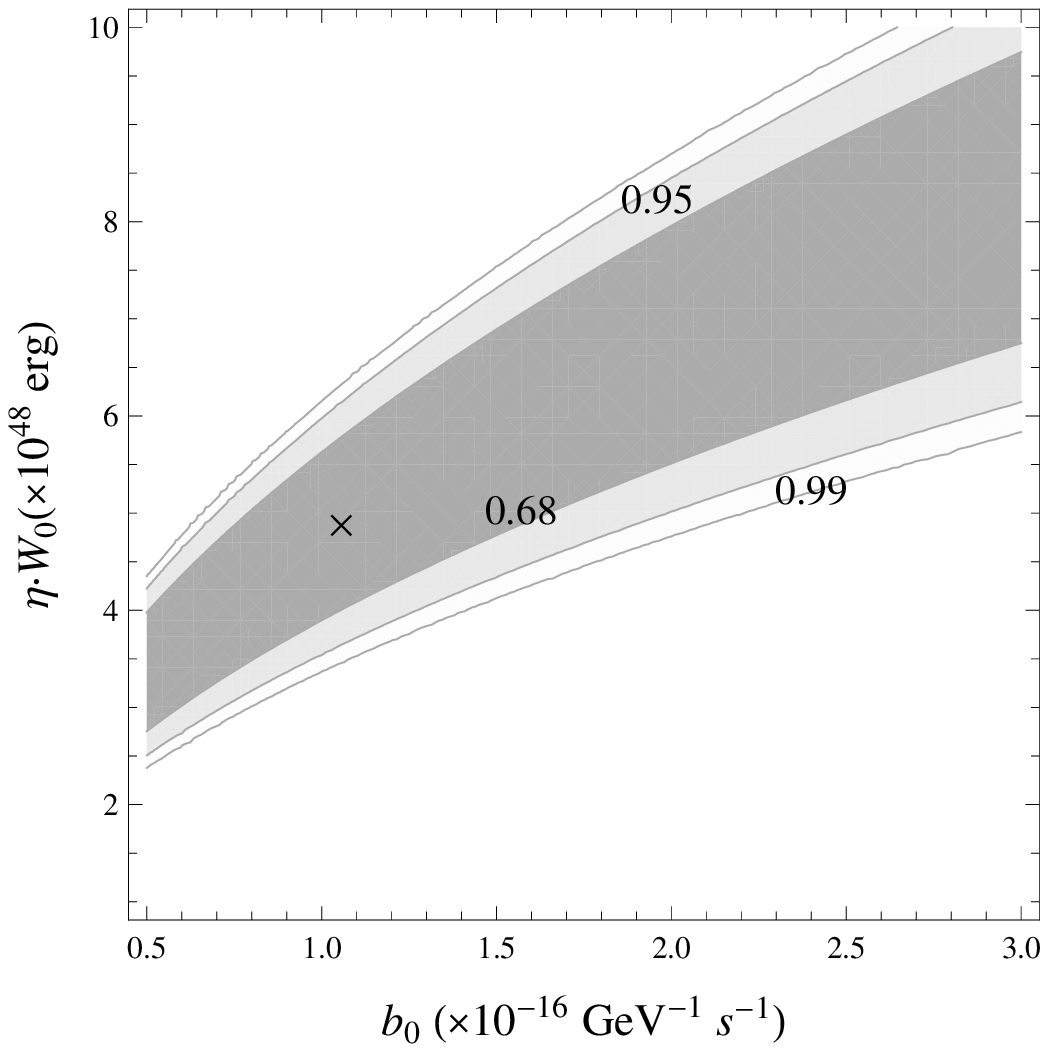}
\vskip 0.15in
\includegraphics[scale = 0.6]{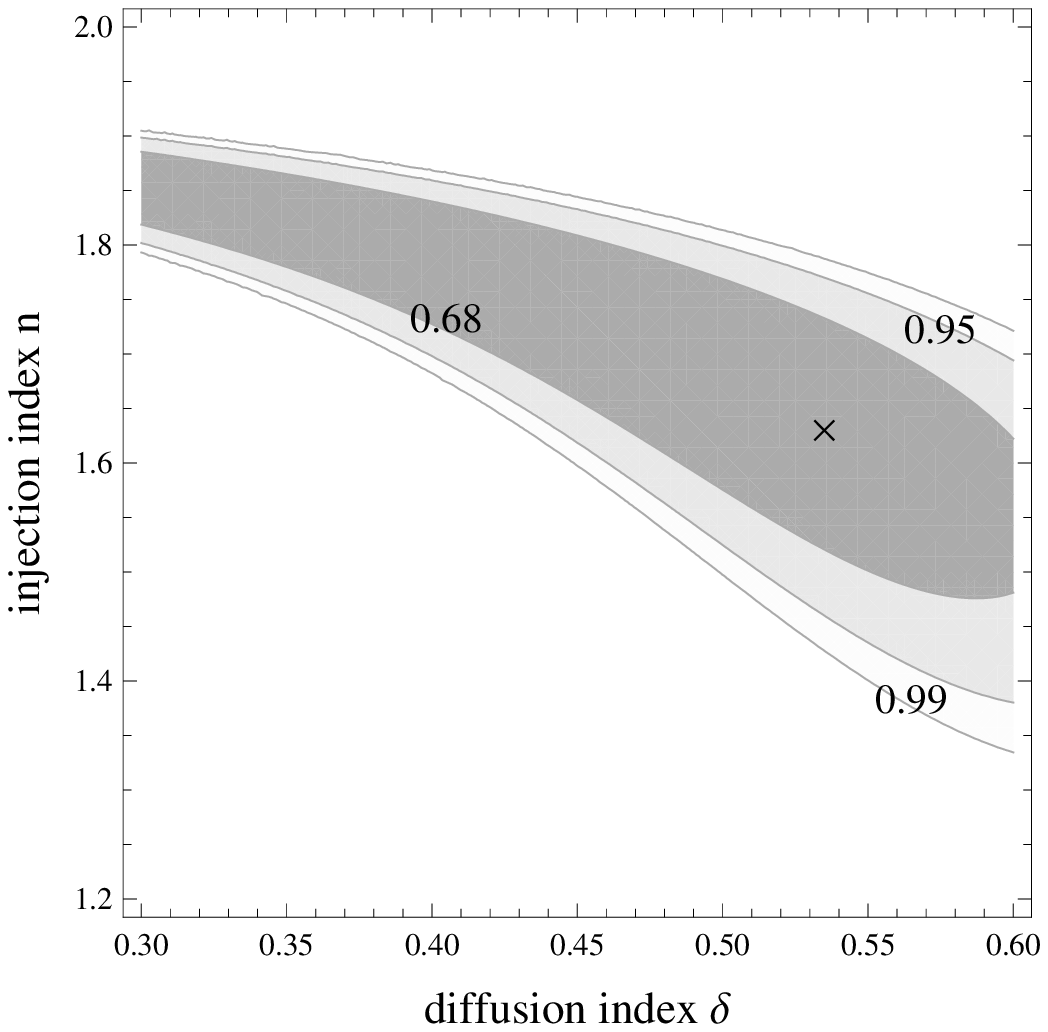}
\hskip 0.15in
\includegraphics[scale = 0.6]{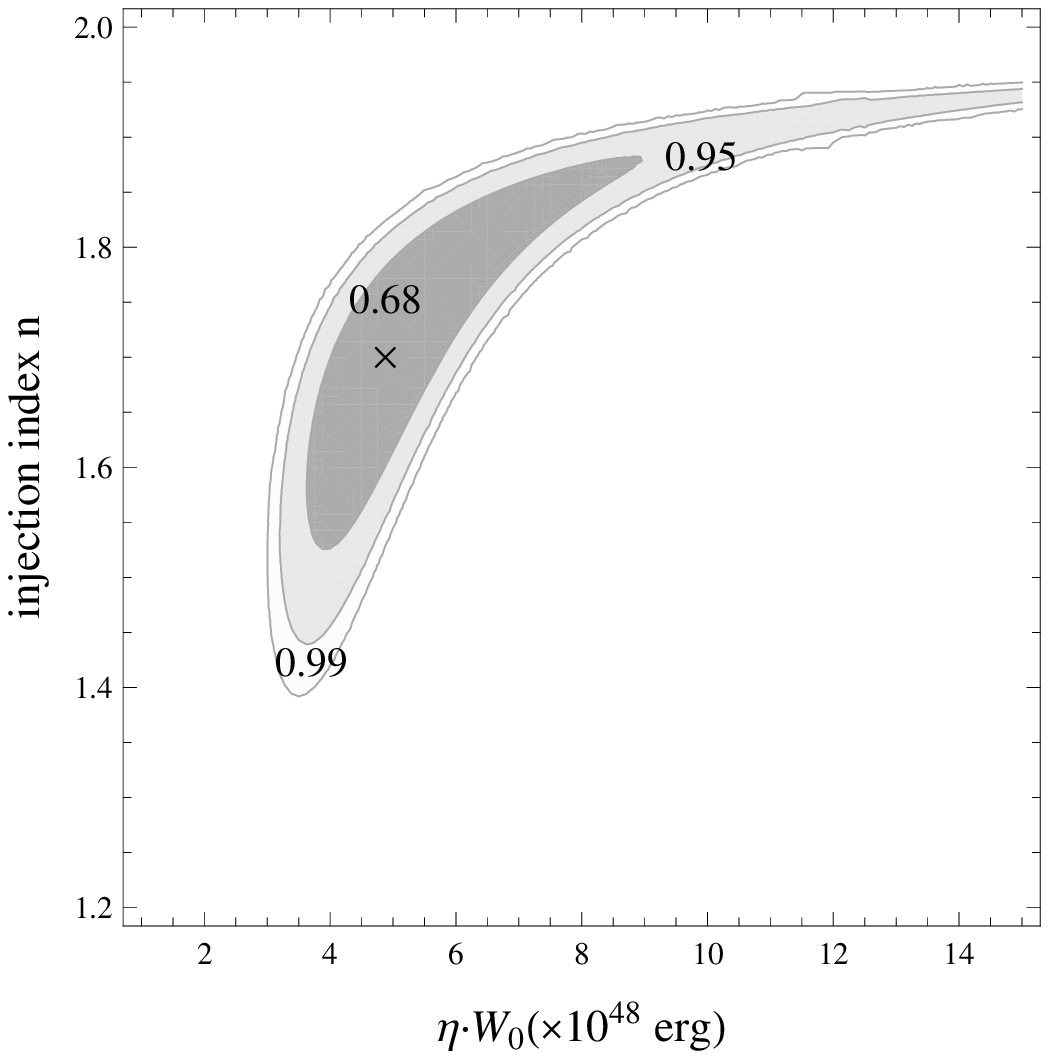}
\end{center}
\caption{\small The fits of the continuous distribution flux to the
ATIC and {\it PAMELA} data.  The best-fit parameters are: $\delta=0.48$,
$b_{0}$=$1.05\times10^{-16}\:\text{GeV}^{-1}\: \text{s}^{-1}$, $\eta                    
W_{0}$=$4.9\times10^{48}$ erg, $n=1.7$, and $M_{\rm stat}=1.0$ TeV.
The contours show the confidence levels relative to the best-fit,
$\sm^2 = (\chi^2 - \chi^2_*) / dof$, where $\chi^2_*$ is the chi
squared for the best-fit parameters.}
\label{ContourPlots}
\end{figure*}

In Fig. 10 
we plot the $68\%$, $95\%$ and $99\%$ CL of
these parameters.  Every contour plot is obtained by varying two
parameters while keeping the rest fixed at their best-fit value.  We
can see that for any value of the energy loss coefficient $b_{0}$
within the region chosen, there is a value of diffusion index $\delta$
that can provide a fit within $95\%$ CL.  Additionally, a higher value
of $n$ suggests a lower value of $\delta$, in agreement with the
calculation of the $e^{+}e^{-}$ flux from a continuous distribution
of pulsars where we expect that $F \sim E^{- n - (\dl + 1) / 2}$ from
Eq.~(\ref{ContDistrScaling}).  Assuming a high value for the injection
index ($n\geq1.6$), the {\it Fermi} and {\it PAMELA} data could be fitted by a
relatively large region of values of the propagation parameters
($0.3\leq \delta \leq 0.6$ and $b_{0}\leq 3\times
10^{-16}\:\text{GeV}^{-1}\: \text{s}^{-1}$).  On the other hand,
values of $n<1.3$ do not seem to give a very good fit to the data, with
any combination of propagation parameters.  Also if the total energy
converted to $e^{+}e^{-}$ is smaller than $2\times 10^{48}$ erg, then
for a pulsar birth rate of 1.8 per kyr regardless of the value of $n$
the {\it Fermi} and {\it PAMELA} data cannot be explained by the continuous
distribution of pulsars.

{\small
\begin{table}[tpb]
\begin{center}
\begin{tabular} {|c|c|c|c|c|}
\hline
Pulsar &$b_{0}$&$\delta$&$n$&$\eta$ \\
Distribution & & & & \\
\hline\hline
Cont. dist. B1 & 0.5-3.0 & 0.30-0.60 & 1.35-1.95 & 0.025-0.11 \\
\hline
Cont. dist. B2 & 0.5-3.0 & 0.30-0.60 & 1.40-1.95 & 0.020-0.085 \\
\hline
4kpc pulsars B1 & 0.5-2.4 & 0.30-0.60 & 1.25-1.80 & 0.025-0.14 \\
\hline
4kpc pulsars B2 & 0.5-2.0 & 0.30-0.60 & 1.15-1.95 & 0.020-0.13 \\
\hline\hline
\end{tabular}
\end{center}
\caption{\small Table of best-fit parameters within $95\%$ CL.  The
values of $b_0$ are in units of 
$ 10^{-16} \:\text{GeV}^{-1}\:\text{s}^{-1}$.  
In the first two lines, the ranges are given for a
continuous distribution of pulsars.  In the last two lines the
equivalent ranges are given for the distribution of all pulsars
within 4 kpc.  B1 and B2 stand for the two backgrounds, simple power
law and more realistic-conventional background \cite{Abdo:2006fq}, respectively.
We assume the mean initial rotational energy $W_{0} =                                      
10^{50}\:\text{erg}$.}
\label{TableCL95}
\end{table}
}

To show the robustness of our procedure we applied the same analysis
for two different backgrounds, a power law background (B1) and a more
conventional background used in \cite{Abdo:2006fq} (B2). 
In Table I, 
we present the $95\% $ C.L. allowed region of
values of the averaged ISM and averaged pulsar properties, using the
two different backgrounds for the continuous pulsar distribution
used. Alternatively, we used the properties of pulsars in the ATNF
database with estimated distances $d<4$ kpc. We present in
Table I 
the derived constraints in the ISM averaged
properties and universal pulsar properties $n$ and $\eta W_{0}$.

Better data on the flux of the high energy $e^{+}e^{-}$ and on the
pulsar birth rate will make this analysis more successful in confining
the parameter space that is relevant for the pulsar scenario. Tighter
constraints of the backgrounds and the parameters of propagation
through the ISM will be needed to confine the properties of pulsars
themselves.




\bibliography{pulsars}         
\bibliographystyle{utphys}   

\end{document}